\documentclass[twocolumn,showpacs,nofootinbib,preprintnumbers,amsmath,amssymb,superscriptaddress]{revtex4-1}

\usepackage{graphicx} 
\usepackage{dcolumn} 
\usepackage{bm} 
\usepackage{color} 
\usepackage{slashed}
\usepackage{dsfont}

\usepackage{hyperref}

\def\Phib{\bar{\Phi}}

\def\Eq#1{Eq.~(\ref{#1})}

\newcommand{\mrm}[1]{\mathrm{#1}}
\newcommand{\mc}[1]{\mathcal{#1}}
\newcommand{\blr}[1]{\left(#1\right)}
\newcommand{\pd}{\partial}
\newcommand{\be}{\begin{equation}}
\newcommand{\ee}{\end{equation}}
\newcommand{\bea}{\begin{eqnarray}}
\newcommand{\eea}{\end{eqnarray}}
\newcommand{\bse}{\begin{subequations}}
\newcommand{\ese}{\end{subequations}}
\newcommand{\sx}{\sigma}
\newcommand{\sy}{\sigma_\mrm{s}}
\def\qqb{\mrm{q\bar{q}}}

\def\Fig#1{Fig.~\ref{#1}}

\def\eq#1{(\ref{#1})}
\def\Eq#1{Eq.~(\ref{#1})}
\newcommand {\apgt} {\ {\raise-.5ex\hbox{$\buildrel>\over\sim$}}\ }
\newcommand {\aplt} {\ {\raise-.5ex\hbox{$\buildrel<\over\sim$}}\ }

\def\s0#1#2{\mbox{\small{$ \frac{#1}{#2} $}}}
\def\0#1#2{\frac{#1}{#2}}

\def\CU{{\mathcal U}}

\newcommand{\Tr}{\mathrm{Tr}}

\graphicspath{{./figures/}}

\begin{document}

\title{Improved Polyakov-loop potential for effective models from functional calculations}

\author{Lisa~M.~Haas} \affiliation{Institut f\"ur Theoretische Physik,
  Universit\"at Heidelberg, Philosophenweg 16, D-69120 Heidelberg,
  Germany} \affiliation{ExtreMe Matter Institute EMMI, GSI,
  Planckstra{\ss}e 1, D-64291 Darmstadt, Germany} \author{Rainer~Stiele}
\affiliation{Institut f\"ur Theoretische Physik, Universit\"at Heidelberg,
  Philosophenweg 16, D-69120 Heidelberg, Germany} \affiliation{ExtreMe
  Matter Institute EMMI, GSI, Planckstra{\ss}e 1, D-64291 Darmstadt,
  Germany} \author{Jens~Braun} \affiliation{Institut f\"ur Kernphysik
  (Theoriezentrum), Technische Universit\"at Darmstadt, Schlo\ss gartenstra\ss e 2, D-64289
  Darmstadt, Germany} \affiliation{ExtreMe Matter Institute EMMI, GSI,
  Planckstra{\ss}e 1, D-64291 Darmstadt, Germany} \author{Jan~M.~Pawlowski} \affiliation{Institut f\"ur
  Theoretische Physik, Universit\"at Heidelberg, Philosophenweg 16,
  D-69120 Heidelberg, Germany} \affiliation{ExtreMe Matter Institute EMMI,
  GSI, Planckstra{\ss}e 1, D-64291 Darmstadt, Germany}
\author{J\"urgen~Schaffner-Bielich} \affiliation{Institut f\"ur Theoretische
  Physik, Universit\"at Heidelberg, Philosophenweg 16, D-69120
  Heidelberg, Germany} \affiliation{ExtreMe Matter Institute EMMI, GSI,
  Planckstra{\ss}e 1, D-64291 Darmstadt, Germany} \affiliation{Institut
  f\"{u}r Theoretische Physik, Goethe-Universit\"at {Frankfurt},
  Max-von-Laue-Stra\ss e 1, D-60438 Frankfurt am Main, Germany}
\date{\today}

\begin{abstract}
  We investigate the quark backreaction on the Polyakov loop and its
  impact on the thermodynamics of quantum chromodynamics. The dynamics of the gluons generating
  the Polyakov-loop potential is altered by the presence of dynamical
  quarks. However, this backreaction of the quarks has not yet been
  taken into account in Polyakov-loop extended model studies. In the
  present work, we show within a 2+1 flavour Polyakov-quark-meson model
  that a quark-improved Polyakov-loop potential leads to a smoother
  transition between the low-temperature hadronic phase and the
  high-temperature quark-gluon plasma phase. In particular, we
  discuss the dependence of our results on the remaining uncertainties
  that are the critical temperature and the parametrisation of the
  Polyakov-loop potential as well as the mass of the $\sigma$-meson.
\end{abstract}



\maketitle

\section{\label{sec:intro}Introduction\protect}
Understanding the properties of the elementary particles in the phase
diagram of strongly correlated matter is one of the major challenges
of present research. Its general structure is dictated by the phase
transition line separating the low-temperature hadronic phase
associated with quark confinement and chiral symmetry breaking and the
chirally symmetric high-temperature quark-gluon plasma phase. There
are several possibilities to investigate this transition under
different conditions. According to the standard model of particle
physics and cosmology, this transition happened at least once and that
is in the early Universe at high temperatures and small net quark
densities (see e.g.~Ref.~\cite{Boyanovsky:2006bf}). Another natural
environment where this transition might take place, now at high densities
and low temperatures, could be supernovae and their remnants, compact
stars (see e.g.~Ref.~\cite{Weber:2004kj}). More detailed information about the
phase diagram can be obtained by large accelerator facilities with
high performance detectors, as the Large Hadron Collider, the
Relativistic Heavy Ion Collider, the Facility for Antiproton and Ion
Research and the Nuclotron-based Ion Collider Facility (see
e.g.~Ref.~\cite{Muller:2012zq}).

Functional continuum methods are well suited for a combined study of
the chiral and confining dynamics of QCD at finite temperature and
density. In recent years, much progress in this direction has been
made within the functional renormalisation group (FRG) approach to
QCD, see Refs.~\cite{Braun:2009gm,Pawlowski:2010ht}, and to
Polyakov-loop extended low-energy effective models, see
Refs.~\cite{Schaefer:2007pw,Skokov:2010wb,Herbst:2010rf,Braun:2011fw,Braun:2012zq,Herbst:2013ail}.
In fact, the latter emerge dynamically from the first-principle
QCD-flow at low energies: the parameters of these models defined below
a hadronic mass scale $\sim 1\,\text{GeV}$ can be extracted from a QCD
RG-flow starting with the classical QCD action at a given high
(perturbative) scale~$\Lambda \gg \Lambda_{\rm QCD}$. For the
Polyakov-loop extended quark-meson model (PQM) this has been discussed in
Refs.~\cite{Schaefer:2007pw,Pawlowski:2010ht,Herbst:2010rf,Herbst:2013ail}. 

Combining first-principle QCD flows with low-energy
effective models can be very fruitful. For example, such an approach
reveals the dynamical connection of confinement and chiral symmetry
breaking from an analysis of the nonperturbative fixed-point structure
of the theory, see Refs.~\cite{Braun:2011fw,Braun:2012zq}. More
generally, the first-principle determination of the input parameters
of such models, as has been described above, can be used to
systematically remove the ambiguities of low-energy effective
models. This systematic procedure utilises the results of the
first-principle flows for the chiral properties, see
Refs.~\cite{Gies:2005as,Braun:2005uj,Braun:2006jd,Braun:2008pi,Braun:2009gm},
and confinement-deconfinement properties, see
Refs.~\cite{Braun:2007bx,Marhauser:2008fz,Braun:2009gm,
  Braun:2010cy,Fister:2013bh}, both in Yang-Mills (YM) theory and QCD at
finite temperature and density. The technical details of this
systematic embedding of low-energy effective models in first-principle
QCD is discussed in Sec.~\ref{sec:FRG}. 

The embedding discussed above also entails that it is even more
interesting to study low-energy effective models on their own in order
to determine the relevant input parameters in these models, i.e.,
those the infrared behaviour of these models is sensitive to. Low-energy
effective models capture the most important properties of
strongly interacting matter and are well suited for basic
investigations and applications, for a recent review see
Ref.~\cite{Fukushima:2013rx}. The most popular ones are the linear sigma
model, also called quark-meson model in this context
\cite{GellMann:1960np, Metzger:1993cu,
  Berges:1997eu,Schaefer:1999em,Lenaghan:2000ey, Scavenius:2000qd,
  Berges:2000ew,Braun:2003ii,Mocsy:2004ab,Schaefer:2004en,Bowman:2008kc, Schaefer:2008hk},
the Nambu--Jona-Lasinio (NJL) model \cite{Nambu:1961tp, Nambu:1961fr,
  Klevansky:1992qe, Buballa:2003qv, Costa:2008yh}, and the Polyakov-loop extended
versions thereof~\cite{Meisinger:1995ih,Pisarski:2000eq,Scavenius:2002ru,
  Mocsy:2003qw, Fukushima:2003fw,Megias:2004hj,
  Ratti:2005jh,Sasaki:2006ww,Roessner:2006xn,Fraga:2007un, Schaefer:2007pw,
  Skokov:2010wb,Schaefer:2009ui, Herbst:2010rf, Schaefer:2011ex, Mintz:2012mz,Herbst:2013ail}. In
these models the gauge part is adjusted to lattice calculations of
 Yang-Mills theory \cite{Scavenius:2002ru,Ratti:2005jh,Roessner:2006xn}. A
first step towards the full inclusion of the backreactions of quarks
on the gauge sector is to estimate the change of the transition
temperature of the Polyakov-loop potential when going from Yang-Mills
theory to QCD~\cite{Schaefer:2007pw}. This can safely be done on the
basis of a perturbative estimate of the change of $\Lambda_{\rm QCD}$
when going from Yang-Mills theory to QCD~\cite{Schaefer:2007pw} which
also has been used for the chiral dynamics
\cite{Braun:2005uj,Braun:2006jd}. However, on the mean field level
such an analysis lacks the full inclusion of glue-matter dynamics to
both the chiral as well as the confinement-deconfinement physics. The
related quantum and thermal fluctuations typically smoothen the
respective phase transitions. Indeed, the mean field models show
sharper transitions in comparison to the full theory, see
Ref.~\cite{Braun:2009gm}. In the present work, we will show that the
functional renormalisation group allows for a comparison of the Yang-Mills
potential with the glue part of the full effective potential in
QCD. We extract the relation of both and apply it then to improve the
Polyakov-loop potential entering effective models, such as
Polyakov-loop extended Nambu--Jona-Lasinio/quark-meson (PNJL/PQM) models.
With this improved Polyakov-loop potential we shall see that the phase
transition becomes smoother and is in better agreement with most
recent results from functional QCD studies and lattice QCD
simulations.

The paper is organised as follows: in the following
section~\ref{sec:FRG} we summarise our results from a functional
renormalisation group approach which we have already used in earlier
studies to compute the Polyakov-loop potential. In particular, we
compare in Sec.~\ref{sec:PQM-other} the pure Yang-Mills potential with
the glue part of the full effective potential in QCD. From this
comparison, we extract a description of how the pure Yang-Mills Polyakov-loop
potential entering PNJL/PQM models has to be modified such that the
quark backreaction on the gluodynamics generating this potential is
effectively taken into account. After we have given a brief summary
of the PQM model in Sec.~\ref{sec:PQM}, we employ our results from
Sec.~\ref{sec:PQM-other} to amend the Polyakov-loop potential entering
PNJL/PQM model studies. In Sec.~\ref{sec:ResultsDis}, we then show how
this improvement affects the temperature dependence of the order
parameters and thermodynamic observables, in particular close to the
phase transition. To complete our studies we discuss the dependence of
our results on the remaining free parameters. Finally, in
Sec.~\ref{sec:DiscOut}, we give our conclusions and outline possible
continuations.

\begin{figure}[b]
  \includegraphics[width=\columnwidth]{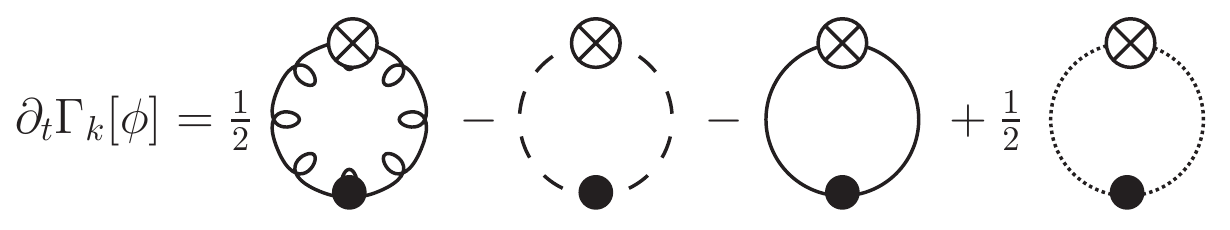}
  \caption{Partially hadronised version of the FRG flow for QCD. The
    loops denote the gluon, ghost, quark and hadronic contributions,
    respectively. The crosses mark the RG regulator term.}
  \label{fig:QCDflow}
\end{figure}
%

\section{\label{sec:FRG} Functional flows in QCD \& low energy
  effective models}

It has been known since long, see e.g.\ 
Refs.~\cite{Berges:2000ew,Gies:2002hq,%
  Pawlowski:2005xe,Gies:2006wv,Schaefer:2007pw,Braun:2008pi,Braun:2009gm,%
Herbst:2010rf,Pawlowski:2010ht,Braun:2011pp}, that low-energy
effective models can be systematically related to full QCD within the
FRG-approach. For Polyakov-loop extended models this follows from the
Landau gauge approach in
Refs.~\cite{Braun:2009gm,Haas:2010bw,Pawlowski:2010ht}, for the Polyakov
gauge see Refs.~\cite{Marhauser:2008fz, Kondo:2010ts}. This setting has been
discussed in details for the PQM model in
Refs.~\cite{Pawlowski:2010ht,Herbst:2010rf, Herbst:2012ht,Herbst:2013ail}. While
chiral symmetry and its dynamical breaking are well described within
these models, confinement is only included in a statistical manner via
a phenomenological Polyakov-loop potential $\cal U$. This potential is
fixed to lattice data of the pure Yang-Mills system at vanishing
chemical potential, hence it is approximately the Polyakov loop
potential in Yang-Mills theory, $V_{\rm YM}$. More specifically, a
Polyakov-loop potential is embedded in these models which reproduces
the temperature dependence of the Polyakov loop and the thermodynamics
of pure Yang-Mills theory as obtained in lattice simulations. As this only
fits two observables which basically do not change below the phase
transition, this leaves us with a big parameter space. Results for the
different potentials have e.g.\ been discussed in
Refs.~\cite{Schaefer:2009ui, Mintz:2012mz}. Also, the coupling of the matter
sector to the gauge sector is lost in such an approach, see also the
discussion in Refs.~\cite{Schaefer:2007pw,Herbst:2010rf,Pawlowski:2010ht,Herbst:2012ht,Herbst:2013ail}.

In summary this entails that the glue potential of full QCD, $V_{\rm
  glue}$, encoding the gauge dynamics in the presence of matter
fields, is replaced by a phenomenological Polyakov-loop potential
$\cal U$. The present work is based on a qualitative improvement of
this approximation towards full QCD by embedding Polyakov loop
extended models within QCD flows.

\subsection{QCD with functional methods}
In this section we summarise results obtained with functional
methods for QCD. The present work mostly relies on the results
obtained in Refs.~\cite{Braun:2009gm,Pawlowski:2010ht} for the QCD phase
diagram. In Ref.~\cite{Braun:2009gm} the FRG approach to the phase diagram
was put forward on the basis of the QCD flow equation in the
background Landau gauge (Landau-DeWitt). It is governed by the flow
for the QCD effective action $\Gamma_k[\bar A;\phi]$,
\begin{eqnarray}\nonumber 
  \partial_t \Gamma_k[\bar A;\phi]&=&\012 \Tr\, G_a[ \bar A;\phi]\,\partial_t R_A
  - \Tr\, G_c[ \bar A;\phi]\,\partial_t R_c\\[1ex]
  & &\hspace{-1cm} -\Tr\, G_q[ \bar A;\phi]\,\partial_t R_q
  + \012 \Tr\, G_H[ \bar A;\phi]\,\partial_t R_H\,,
\label{eq:QCDflow}\end{eqnarray}
depicted in \Fig{fig:QCDflow}.
Here, $\bar A$ is the gluonic background and $\phi=(a\,,\,c\,,\,\bar c
\,,\,q\,,\,\bar q\,,\,{\rm hadrons})$ are the dynamical fluctuations
with $A=\bar A+a$. The flow \eq{eq:QCDflow} of the effective action or
free energy only depends on the full field-dependent fluctuation
propagators $G_\phi[\bar A;\phi]$,
\pagebreak
\begin{eqnarray}\label{eq:G}
  G_{\phi}[\bar A;\phi](p,q)=\left(\0{1}{\Gamma_{k}^{(2)}[\phi]
      +R_k}\right)_{\phi \phi }(p,q)\,, 
\end{eqnarray} 
and the regulator functions $R_\phi$. The regulators $R_\phi$ depend
on background covariant momenta $\bar D=\partial- i\,g\, \bar A$ such
that for small momenta it acts as a mass and for large momenta it
vanishes sufficiently fast; i.e., $R_k$ acts as a momentum-dependent
additional mass term. The present approach has the advantage that
hadronic states can be included successively within dynamical
hadronisation, see Refs.~\cite{Gies:2001nw,Gies:2002hq,%
  Pawlowski:2005xe,Floerchinger:2009uf}. This simply entails that part
of the quark and gluonic fluctuations are treated separately as they
carry hadronic resonances. It is worth emphasising that this is done
dynamically within the QCD-setting so there is no danger of
overcounting terms present within an effective model setup. The
QCD results obtained in the present work are computed within the
approximation to two-flavour QCD detailed in
Refs.~\cite{Braun:2009gm,Pawlowski:2010ht} which explicitly takes into
account the lowest mesonic multiplet, the $\sigma$-meson and the
pions, $\sigma,\vec \pi$.

The FRG-approach to QCD outlined above as well as other functional
approaches allow us to access the physics of the
confinement-deconfinement phase transition in a simple way, see
Refs.~\cite{Braun:2007bx,Braun:2009gm,Braun:2010cy,Pawlowski:2010ht,Fister:2013bh}. 
The symmetry behind this transition, the center symmetry, is only an
unbroken symmetry for infinitely heavy quarks. Then, QCD reduces to a
pure SU$(N_c)$ gauge theory. The commonly used order parameter is the 
expectation value of the Polyakov-loop operator, 
\begin{eqnarray}
&& \Phi\blr{\vec{x}}\equiv \Phi[A(\vec{x})] \nonumber\\
&& \qquad= \frac{1}{N_c} \text{tr}_{\rm c} \left[ {\mathcal P} \exp \blr{ i 
\bar{g}\int_0^\beta dx_0 \,A_0\blr{x_0,\vec{x}}} \right]\;,
\end{eqnarray}
where $\beta=1/T$ is the inverse temperature, $\bar{g}$~denotes the
bare gauge coupling and $\mathcal P$ stands for path ordering. The
Polyakov loop is then given by~$\langle \Phi\blr{\vec{x}} \rangle$.

Strictly speaking, the Polyakov loop $\langle
\Phi\blr{\vec{x}}\rangle$ is an order parameter for center symmetry
breaking, see e.g.~Ref.~\cite{Greensite:2003bk}. However, its
logarithm can be associated to half of the free
energy~$F_{q\bar{q}}$ of a quark-antiquark pair at infinite
distance. A center-symmetric confining phase is indicated by a
vanishing Polya\-kov loop and implies that the free energy of a static
quark is infinite. On the other hand, the deconfined phase is
associated with a finite free energy of a static quark and, in turn, a
finite Polyakov loop. The latter implies that center symmetry is
broken spontaneously at high temperatures in Yang-Mills theories. In
the presence of (light) dynamical quarks, the Polyakov loop is then
finite for all temperatures but still becomes small in the ``confined"
low-temperature phase. This can be eventually traced back to the fact
that quarks explicitly break the (global) center symmetry of the
theory.

In addition to the standard Polyakov loop, other order parameters
for quark confinement have been introduced. In this work we shall 
consider an order parameter which is closely related to the Polyakov
loop, namely~$ \Phi[\langle A_0\rangle]$. In Polyakov-Landau-DeWitt
gauge, it can be indeed shown that $ \Phi[\langle A_0\rangle] $ also
is an order parameter, see
Refs.~\cite{Braun:2007bx,Marhauser:2008fz}. Here, $\langle A_0\rangle$
is a constant element of the Cartan subalgebra of the gauge group and
denotes the ground state of the order-parameter potential which is the
Polyakov-loop potential. Perturbatively, the related effective
potential $V[\langle A_0\rangle ]$ has first been computed in
Refs.~\cite{Weiss:1980rj,Gross:1980br}.

Based on a functional RG approach, first nonperturbative QCD studies
of this potential, including a computation of the phase transition
temperatures for SU($N_{\rm c}$) Yang-Mills theories, have been
presented in
Refs.~\cite{Braun:2007bx,Marhauser:2008fz,Braun:2010cy}. In
Ref.~\cite{Fister:2013bh} the computation has been extended to general
functional methods, i.e., Dyson-Schwinger equations and the
2PI-approach using the thermal propagators obtained in
Refs.~\cite{Fister:2011uw,Fister:2011um}. In
Refs.~\cite{Braun:2009gm,Pawlowski:2010ht} the approach has been applied to
$N_f=2$ QCD in the chiral limit. This computation within two-flavour
QCD includes the full backcoupling of the matter sector on the
propagators of the gauge degrees of freedom via dynamical quark-gluon
interactions~\cite{Gies:2002hq,Braun:2006jd,Braun:2008pi}. In
particular, this includes the resolution of the full momentum
dependence of the ghost and gluon
propagators~\cite{Ellwanger:1995qf,Bergerhoff:1997cv,Pawlowski:2003hq,%
  Fischer:2004uk,Fischer:2008uz,Fister:2011uw,Fister:2011um}. Recently,
the nonperturbative Polyakov-loop potential has also been studied using
various other
approaches~\cite{Dumitru:2012fw,Diakonov:2012dx,Sasaki:2012bi,%
  Ruggieri:2012ny,Fukushima:2012qa,Reinhardt:2012qe,Kashiwa:2012td}.

In the present work, we refrain from repeating the details of the
above works but only refer to it whenever it is required. For general
QCD-related introductions and reviews to our functional RG approach,
we refer the reader to
Refs.~\cite{Litim:1998nf,Berges:2000ew,Pawlowski:2005xe,Gies:2006wv,%
  Schaefer:2006sr,Pawlowski:2010ht,Braun:2011pp}.

\subsection{From QCD to Polyakov-loop extended chiral models} 
\label{sec:PQM-other}

\begin{figure}
  \includegraphics[width=.6\columnwidth]{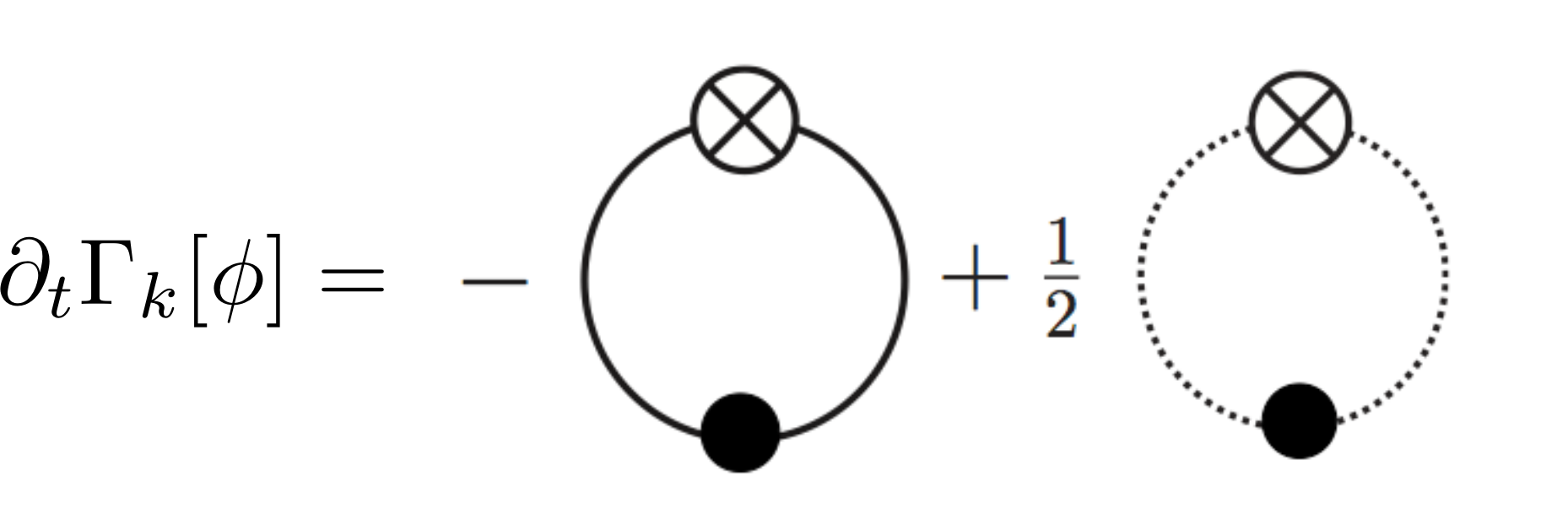}
  \caption{FRG flow for the matter sector of QCD. The
    loops denote the quark and hadronic contributions,
    respectively. The crosses mark the RG regulator term.}
  \label{fig:QHflow}
\end{figure}

For temperatures below the chiral and confinement-deconfinement phase
transitions the glue dynamics decouples from the matter dynamics. In
the Landau gauge this physical decoupling is realised simply by a
mass gap in the gluon propagator, that is $ p^2 G_a(p^2)\to 0$ for
small momenta, see e.g.~Ref.~\cite{Fischer:2008uz} and references
therein. As the ghost-matter coupling is mediated by the gluon this
decoupling extends to the full ghost-gluon dynamics. Hence the pure
glue sector of the theory decouples as expected, and the QCD flow
equation reduces to 
\begin{equation}
  \partial_t \Gamma_k[\bar A;\phi]=-\Tr\, G_q[\bar A;\phi]\,\partial_t R_q
  + \012 \Tr\, G_H[\bar A;\phi]\,\partial_t R_H\,,
\label{eq:QHflow}\end{equation}
which is depicted in \Fig{fig:QHflow}.
The background gluon field as
well as the ghost and gluon fluctuation fields $A$ and $c,\bar c$ are
simply spectators in \Eq{eq:QHflow}. If setting the spectator fields
to zero the flow equation \eq{eq:QHflow} reduces to that of generic
low-energy models of QCD. The respective models are then singled out
by specifying the hadronic content of the flow. In
Eq.~\eqref{eq:QHflow}, the presence of a quark loop should not be
confused with the presence of quarks as observable (asymptotic)
states. The presence of this loop only states that quark loops are
still present ``internally" and required to give a microscopic
description of the interaction of the hadronic degrees of freedom. In
the same spirit, gluon loops also play a role in the low-energy sector
even though they do not represent observable (asymptotic) states in
this regime as well. However, gluon loops govern the dynamics of the
confinement order-parameter potential which in turn determines the
background gluon field entering~Eq.~\eqref{eq:QHflow}, see also our
discussion below.

The most prominent example are NJL-type models. In these models the
hadronic loops are missing completely and the full matter dynamics is
described by the quark loop. This is in correspondence to a
description of the low-energy matter dynamics of QCD purely in terms
of quark correlation functions. This only captures the correct
QCD-dynamics if the hadronic spectrum is taken into account via
resonances in the scattering amplitudes of quarks, e.g.\ in the
quark-quark scattering kernel. The latter is directly related to the
four point function $\Gamma^{(4)}$ of two quark-antiquark pairs. It
is worth noting in this context that the flow equation \eq{eq:QHflow}
includes vertex and propagator corrections to the theory and hence
naturally incorporates the so-called nonlocal NJL
models~\cite{Hell:2008cc,Kondo:2010ts,Kashiwa:2011td}. For a
detailed discussion of the presence and emergence of nonlocal
fermionic vertices in RG flows, we refer the reader to
Ref.~\cite{Braun:2011pp}. 

Even though this incorporates in principle the full dynamics, hadronic
contributions are more easily taken into account via (dynamical)
hadronisation, see
Refs.~\cite{Gies:2001nw,Gies:2002hq,Pawlowski:2005xe,%
  Braun:2008pi,Floerchinger:2009uf}. This technique describes the
following physical mechanism: when lowering the RG-scale in the QCD
flow equation \eq{eq:QCDflow}, \Fig{fig:QCDflow}, strongly-bound quark
correlations e.g.\ quark-antiquark (mesons), diquark, and
three-quark (baryons) correlations, will be dynamically created by the
flow. These correlations can be parameterised in terms of mesonic,
diquark and hadronic operators respectively. Note that the diquark
operators do not describe asymptotic states but intermediate
correlations. The respective coupling of the hadronic operators is
dynamically enhanced by the quark-gluon fluctuations at higher scales,
while it takes over the dynamics from the quark-gluon sector at lower
scales, hence the name dynamical hadronisation. This
fluctuation-triggered hadronisation within the QCD-flow clearly avoids
any double-counting problem: the hadrons are created dynamically, the
respective couplings are created from the QCD input which only depends
on a single input parameter, the strong coupling $\alpha_s$.

\pagebreak
Quantum
and thermal fluctuations are self-consistently taken into
account. Still, at low scales, say for $k\lesssim 1\,{\rm GeV}$, the
pure glue sector decouples and we are left with \Eq{eq:QHflow}. Note
that within such a procedure the initial effective action
$\Gamma_\Lambda$ at $\Lambda \approx 1\,{\rm GeV}$ serves as the
classical action of the quark-hadronic low-energy model and its
coupling constants are directly derived from QCD.

\begin{figure}[b]
  \includegraphics[width=.55\columnwidth]{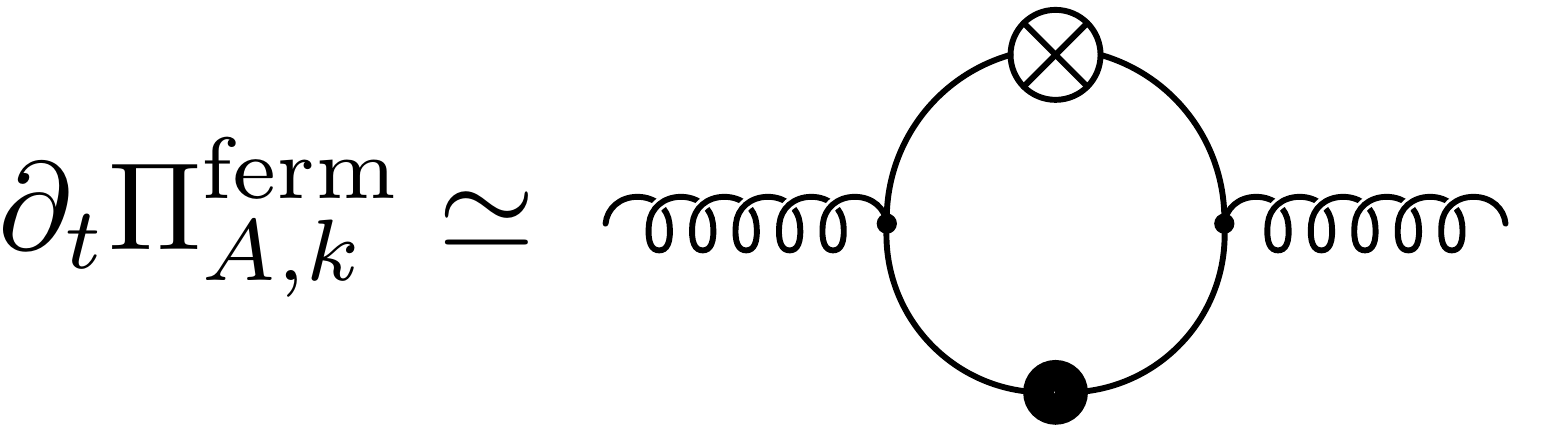}
  \caption{Quark polarisation contribution to the gluon propagator
  representing a contribution to the matter backcoupling.}
  \label{fig:quark_loop}
\end{figure}

In the present work we consider such a quark-hadronic model at finite
temperature and vanishing density: close to the phase boundary between
the quark-gluon plasma phase and the hadronic phase and at not too
large chemical potential mesonic degrees of freedom, in particular the
pion and $\sigma$-fluctuations, become important. Here, we assume that
the associated bosonic fields can be described as composites of
fermions and do not carry an internal charge, such as colour or flavour:
$\sigma \sim \bar{q}q$ and~$\vec{\pi} \sim \bar{q}\gamma_5 \tau q$. A
priori, these fields are just auxiliary fields introduced by means of
a Hubbard-Stratonovich transformation of a purely fermionic
theory. Since these composite objects have considerable overlap with
the full meson operators, they are well suited to describe the meson
dynamics in a first approximation. Keeping this in mind, the
(effective) action of QCD can be expanded in powers of these mesonic
operators, at least for small momentum scales. Momentum dependencies
and corrections of higher order can conveniently and systematically be
taken into account, e.g., by means of a derivative expansion of the
effective action, see Ref.~\cite{Braun:2011pp} for a review. In the
fully coupled QCD RG-flow, the strength and the momentum dependence of
the various couplings can then be derived from the microscopic QCD
action. In fact, the loops depicted in Fig.~\ref{fig:QCDflow} are
intrinsically coupled to each other. In particular, there are
contributions from the matter sector to the diagrams for the gluon
propagator, i.e., the quark part of the gluonic vacuum polarisation,
see Fig.~\ref{fig:quark_loop}.

Hence we only take into account mesonic correlations in terms of
effective operators. Baryonic fluctuations are considered as
subleading at vanishing density and are partially taken into account
in the quark correlations. Moreover, the gluonic background is
determined by the constant solution of the QCD equations of
motion. Such a solution can always be rotated in the Cartan,
\begin{equation}\label{eq:A0const}
\langle A_\mu\rangle =\delta_{\mu 0} \langle A^3_0\rangle \tau^3 
+\langle A^8_0\rangle \tau^8\,,
\end{equation}
with constant $\langle A_0^{3,8}\rangle$. The minima and maxima of
the potential are accessed for $\langle A_0^8\rangle=0$, see
Refs.~\cite{Braun:2007bx,Marhauser:2008fz,Braun:2009gm,%
Braun:2010cy,Fister:2013bh}. Then
the Polyakov loop reads
\begin{eqnarray}\label{eq:Polloop}
  \Phi[A_0]=\frac{1}{3}\left[1+2 \cos \blr{\frac{1}{2}\beta g A_0}\right]\,,\hskip1ex {\rm with} 
  \hskip1ex A_0= \langle A^3_0\rangle\,.
\end{eqnarray}
In summary this leads to
quark-meson flows \eq{eq:QHflow} in constant temporal gauge field
backgrounds which define the PQM
model \cite{Schaefer:2007pw}, including its matter quantum and thermal
fluctuations. We note that for the NJL-type model this defines the
PNJL model including its quantum and
thermal fluctuations. The generalised version is the Polyakov-loop
extended quark-hadron (PQH) model.

\begin{figure}
\includegraphics[width=.6\columnwidth]{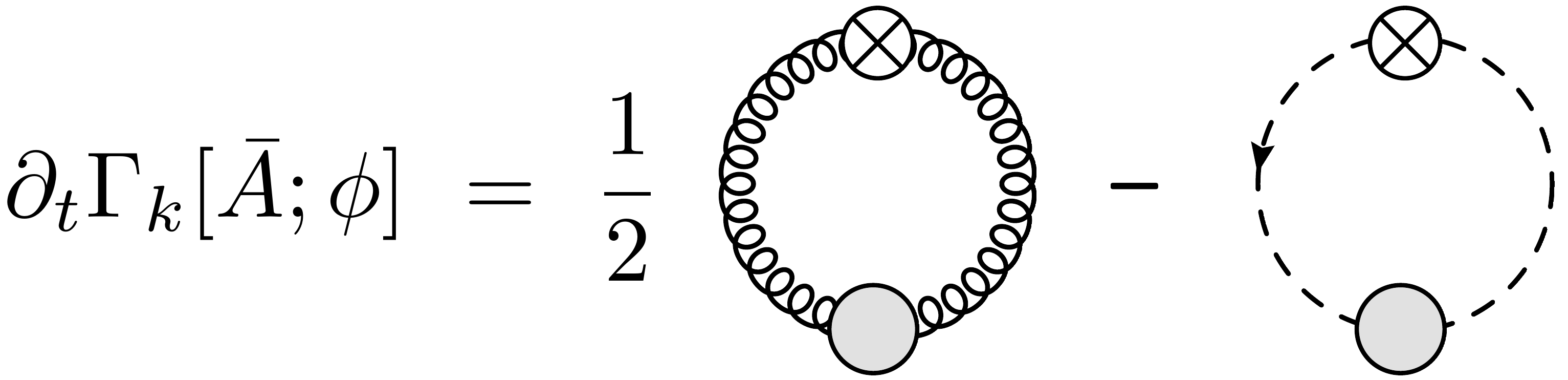}
\caption{Functional flow for the effective action. Lines with filled
  circles denote fully dressed field-dependent propagators
  \eq{eq:G}. Crossed circles denote the regulator insertion
  $\partial_t R_k$. }
\label{fig:YMflow}
\end{figure}

Even though the glue-dynamics decouples for low temperatures, the gluonic 
background $\bar A=\langle A\rangle$ does play an important r$\hat{\rm
  o}$le. In the confining phase it screens the propagation of the
quarks which confines them even statistically, leave aside the
confining potential. The expectation value $ \langle A\rangle$ is
given by the solution of the QCD equations of motion which also
requires the first part of the QCD flow, that is
\begin{equation}
  \partial_t \Gamma_k[\bar A;\phi]=\012 \Tr\, G_a[ \bar A;\phi]
  \,\partial_t R_A
  - \Tr\, G_c[ \bar A;\phi]\,\partial_t R_c\,. 
\label{eq:YMflow}\end{equation}
depicted in \Fig{fig:YMflow}.
%
%
The fluctuation $\phi$ is evaluated at
the equation of motion (EoM), $\phi=\bar\phi$: the gluonic fluctuation
background is vanishing, $\bar a=0$, whereas the mesonic background is
given by $\sigma=\bar\sigma$ and $\vec \pi=0$ leading to the running
quark masses in the quark vacuum polarisation depicted in
\Fig{fig:quark_loop}. This leads us finally to
\begin{equation}\label{eq:Vglue} 
  \partial_t V_{\rm glue}[A_0]= \0{1}{\beta\, {\cal V}} \partial_t 
\Gamma_k[A_0;\bar \phi]\,,
\end{equation}
for constant backgrounds $A_0$ and spatial volume ${\cal
  V}$. Structurally, \Eq{eq:YMflow} resembles the pure Yang-Mills flow
equation, 
\begin{equation}\label{eq:VYM} 
  \partial_t V_{\rm YM}[A_0]= \0{1}{\beta\, {\cal V}} \partial_t 
\Gamma_{{\rm YM},k}[A_0;a=0,C=0,\bar C=0]\,,
\end{equation}
which also has the pictorial form of \Fig{fig:YMflow}. The ghost and
gluon propagators in \Eq{eq:Vglue}, however, are those of QCD. In
particular, the flow of the gluon propagator receives contributions
from matter loops, e.g.\ the quark contribution to the vacuum
polarisation, see \Fig{fig:quark_loop}.

It has been already mentioned at the beginning of this section that in
the standard approach to Polyakov-loop extended models the glue
potential $V_{\rm glue}$ in \Eq{eq:Vglue} is approximated by an
approximation to its Yang-Mills analogue, $V_{\rm YM}$ in
\Eq{eq:VYM}. To bring these studies closer to (full) QCD, however,
this Polyakov-loop potential effectively generated by pure
gluodynamics has to be replaced by the QCD glue potential, i.e., by
the contribution stemming from the gauge degrees of freedom in the
presence of dynamical quarks. It is therefore beneficial to amend
these model calculations by utilising the available information on
this glue part of the potential.

In the present work, we only consider
Polyakov-loop potentials which have been computed by considering two
quark flavours with vanishing current masses. The strange quark has
been ignored. We expect that the corrections resulting from the
inclusion of a strange quark are subleading in our present studies.
Following
Refs.~\cite{Braun:2007bx,Marhauser:2008fz,Braun:2009gm,Braun:2010cy,Fister:2013bh},
we have computed the nonperturbative Polyakov-loop potential using the
above mentioned functional RG approach taking into account the
backreaction of the quark degrees of freedom on the gluon
propagators\cite{Braun:2009gm}.

\begin{figure}
	\centering
	\includegraphics[width=0.47\textwidth]{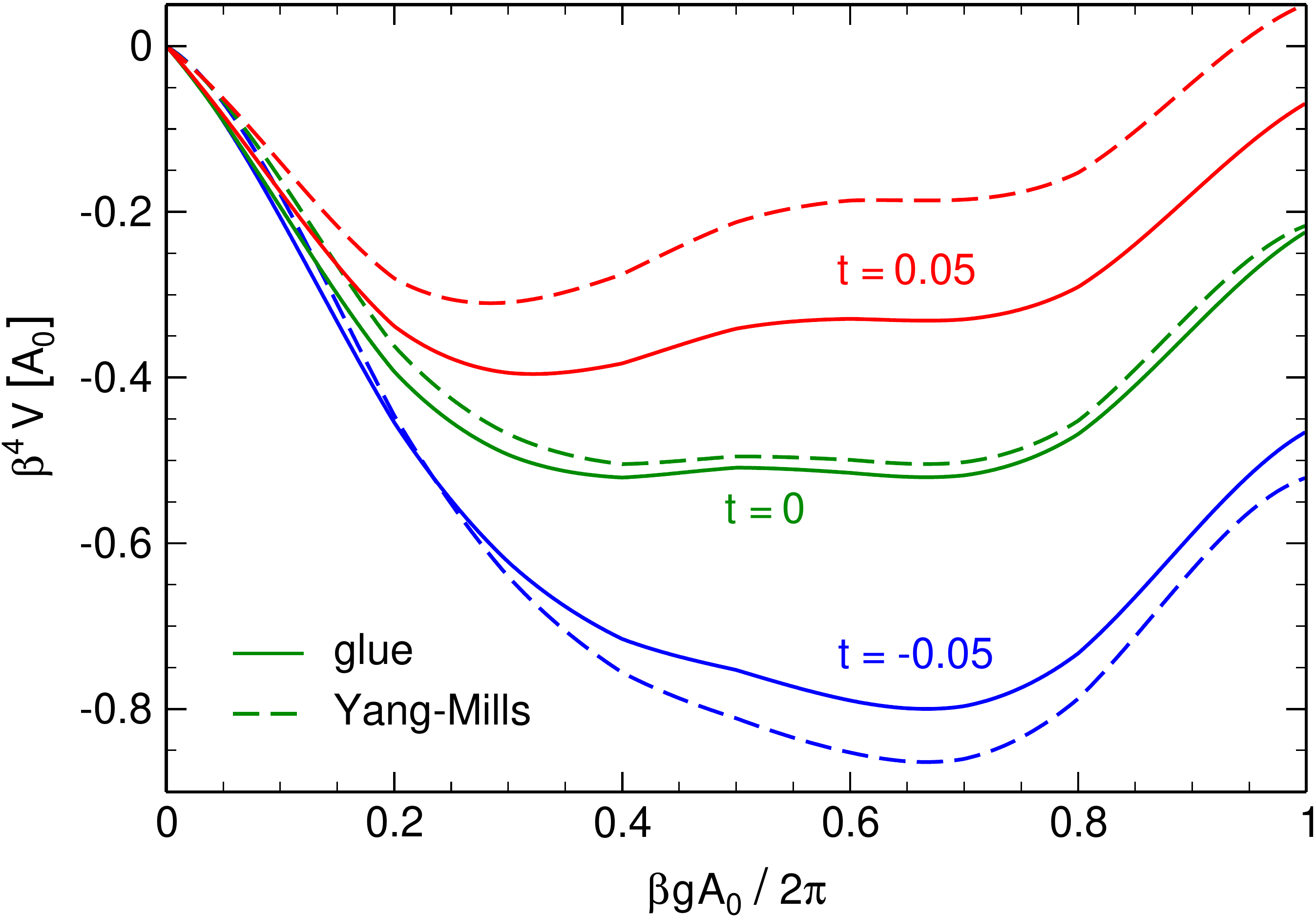}
	\caption[]{Comparison of the SU$(3)$ Yang-Mills and glue effective
          potentials as functions of the background gauge field
          $ A_0=\langle A_0^3\rangle$ for various reduced temperatures, see the
          text for the definition of the reduced temperatures. The
          form of the potentials is very similar, however, the
          temperature scale changes. Only the glue part of the full
          effective potential is shown here and compared to the Yang-Mills
          potential.}
	\label{fig:GlueYMPotentials}
\end{figure}

\begin{figure}
	\centering
	\includegraphics[width=0.47\textwidth]{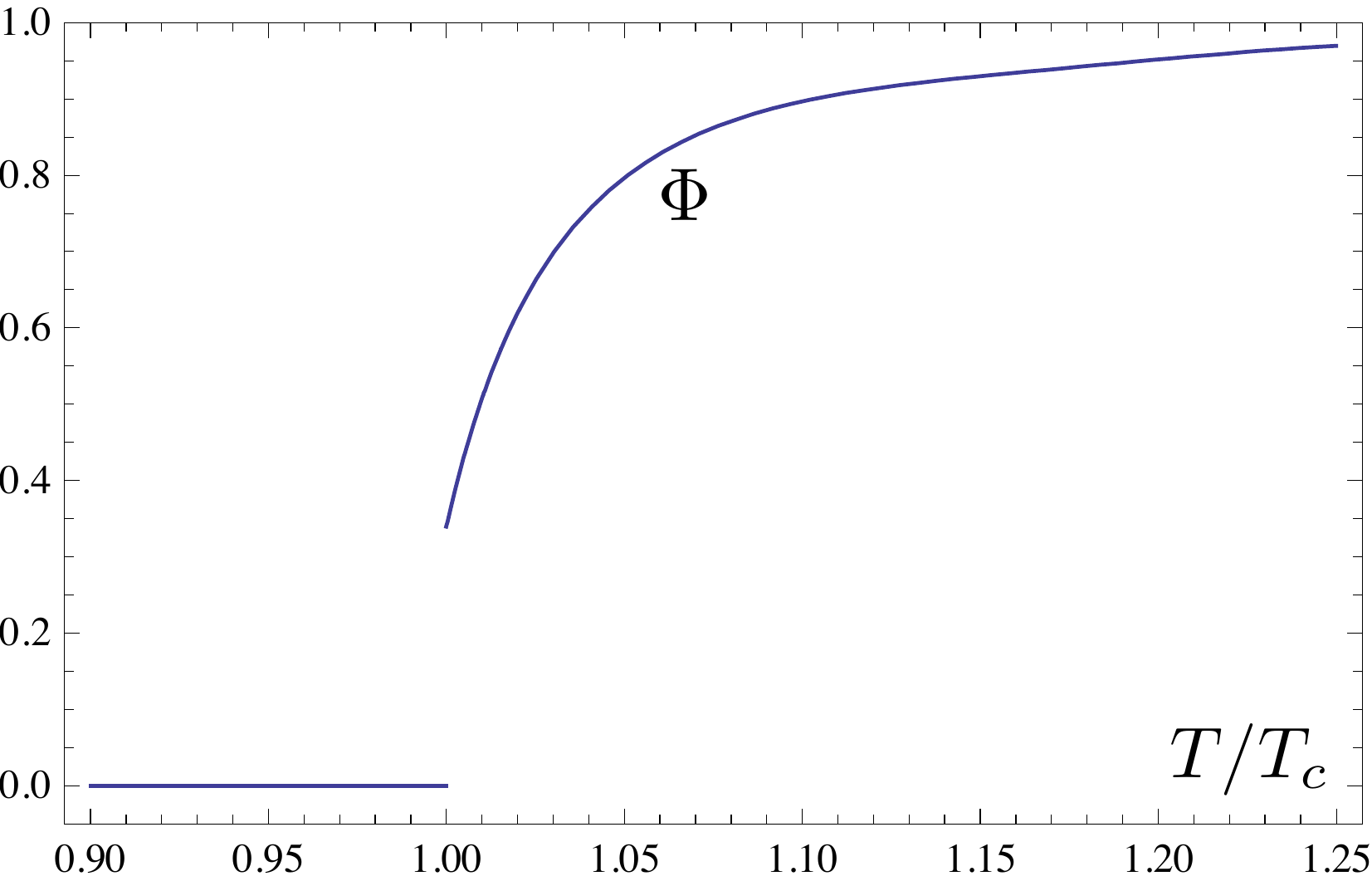}
	\caption[]{Yang-Mills order-parameter $\Phi[\langle
          A_0\rangle]$ as a function of the normalised temperature
          $T/T_c$ with $T_c=276 {\rm MeV}$.}
	\label{fig:SU3PolFRG}
\end{figure}

In Fig.~\ref{fig:GlueYMPotentials},
we compare the effective potential of SU$(3)$ Yang-Mills theory as obtained in
Refs.~\cite{Braun:2007bx,Braun:2010cy,Fister:2013bh} with the
Polyakov-loop potential of the glue sector of our present study.
%
%
The latter includes also the quark part of the gluonic vacuum polarisation
but does not include the fermionic part of the full potential. The
related order-parameters $\Phi[\langle A_0\rangle]$ derive by
inserting the temperature-dependent gauge field value at the minimum into
\Eq{eq:Polloop}. The order parameter shows a first-order phase
transition for both, the glue potential and the Yang-Mills
potential. It is displayed for Yang-Mills theory in
\Fig{fig:SU3PolFRG}.
%
%
We observe immediately in
\Fig{fig:GlueYMPotentials} that the temperature scale of the various
Yang-Mills potentials and the glue potentials differ. The differences are induced by the
matter fluctuations altering the propagators of the gauge fields. This
has already been anticipated in Ref.~\cite{Schaefer:2007pw}, where a
phenomenological hard-thermal-loop (HTL) estimate led to $T^\mrm{HTL}_{\text{cr}}=208$~MeV
for two quark flavours.

\begin{figure}
	\centering
	\includegraphics[width=0.47\textwidth,]{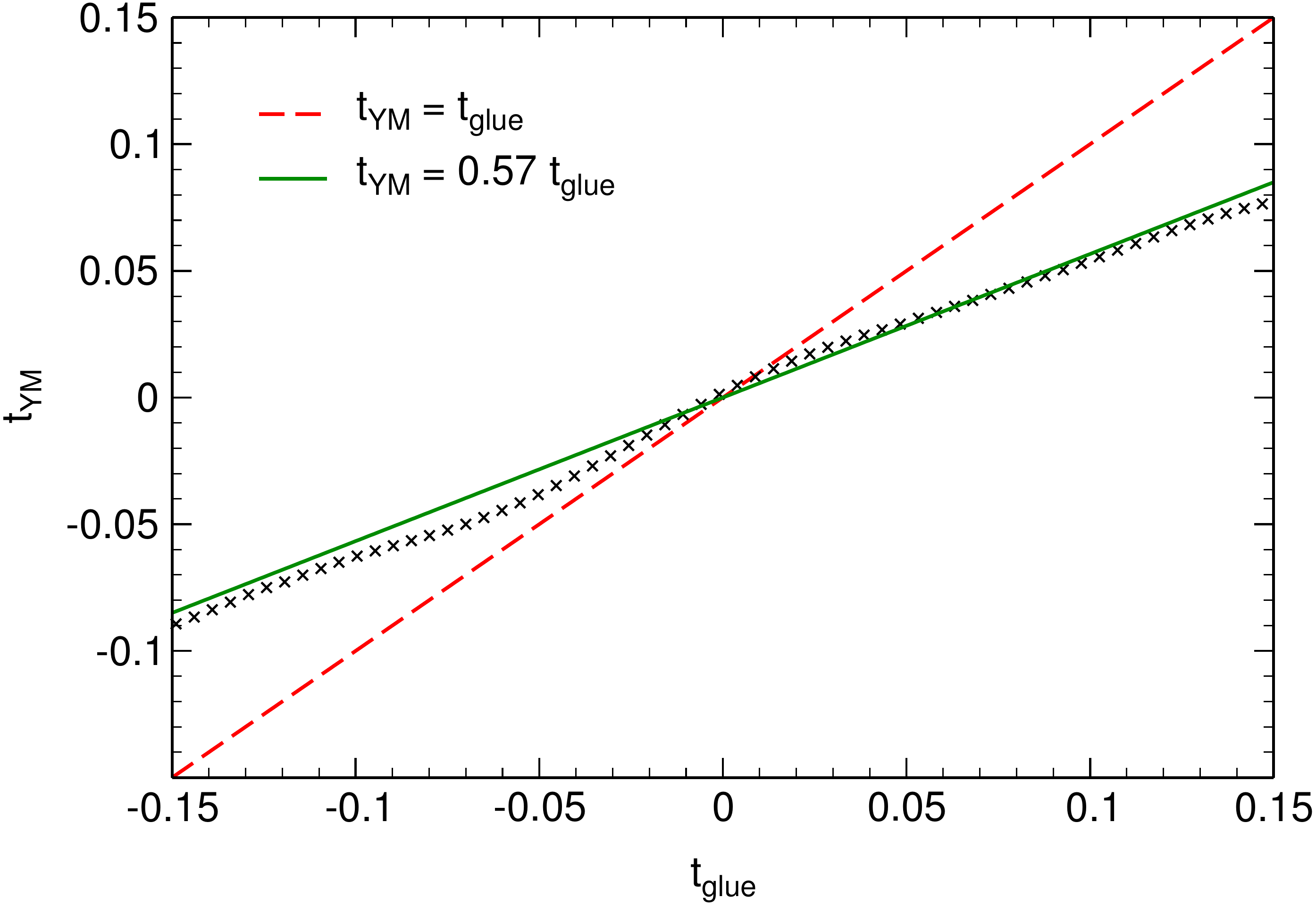}
	\caption[]{Relation between the two temperature scales of pure
          Yang-Mills theory and the glue part of our present study including
          the backreaction of the quarks on the gauge fields. The solid
          (green) line is our fit to the numerical data (black crosses)
          in the range around the transition temperatures. A clear
          deviation from linear behaviour with slope one (dashed red line),
          i.e., data from pure Yang-Mills theory, can be recognised.}
	\label{fig:tYM_tglue}
\end{figure}

\pagebreak
Although the temperature scales are different, the shape of the
order-parameter potential~$V$ is not, see
Fig.~\ref{fig:GlueYMPotentials}. We can exploit this observation to
estimate how the temperature of a given pure Yang-Mills Polyakov-loop
potential has to be modified to be closer to the QCD potential. To
this end, we need to define a measure for the comparison of the
potentials. We use
\begin{equation}
\int_0^{\frac{2\pi T}{\bar{g}}}  d A_0\, \left |V_{\rm YM}( A_0) 
-V_{\rm glue}(A_0)\right|^2\,.
\label{eq:measure}
\end{equation}
Moreover, we introduce the reduced temperatures
\begin{equation}
t_{\text{glue}}=\frac{T-T^{\text{glue}}_{\text{cr}}}{T^{\text{glue}}_{\text{cr}}},
\quad t_{\text{YM}}=\frac{T-T^{\text{YM}}_{\text{cr}}}{T^{\text{YM}}_{\text{cr}}},
\end{equation}
where $T^{\text{glue}}_{\text{cr}}=203\,\text{MeV}$. Note that this
is not the critical temperature obtained from the full effective
potential including pure quark loops but only from the contribution
stemming from the glue part of the potential for two massless quark
flavours \cite{Braun:2009gm}. We add that the absolute scale in
Ref.~\cite{Braun:2009gm} was not computed in a chiral extrapolation of
the theory with physical quark masses. Hence, in the applications to
the PQM model we will leave the glue critical temperature as a free
parameter, maximally ranging from
\begin{equation}\label{eq:Tabsolute}
180\,{\rm MeV} \lesssim
  T^{\text{glue}}_{\text{cr}} \lesssim 270\,{\rm MeV} \,.
\end{equation}
%
\begin{table}[b]
  \caption{Critical temperature of the glue effective potential 
    $T^{\text{glue}}_{\text{cr}}$ for $N_f$ massless flavours and in case 
    of an additional massive quark flavour with the {{current}} strange 
    quark mass of 95\,MeV \cite{Beringer:1900zz}, according to Ref.~\cite{Schaefer:2007pw}.}
  \begin{ruledtabular}
  \begin{tabular}{lccccc}
    $N_f$ & 0 & 1 & 2 & 3 & $2+1$ \\ \hline 
    $T^{\text{glue}}_{\text{cr}}$ [MeV] & 270 & 239 & 208 & 178 & 182 \\
  \end{tabular}
  \end{ruledtabular}
  \label{tab:T0_Nf}
\end{table}
%
The upper limit in \Eq{eq:Tabsolute} is the critical temperature of
Yang-Mills theory, the lower limit is the estimate in
Ref.~\cite{Schaefer:2007pw}, see also Table~\ref{tab:T0_Nf}. 
For
$T^{\text{YM}}_{\text{cr}}$, on the other hand, we find
$T^{\text{YM}}_{\text{cr}}=276$~MeV~\cite{Braun:2007bx,Braun:2010cy,Fister:2013bh}
which is in quantitative agreement with the results from lattice
studies. The comparison of the potentials then yields the translation
of the two temperature scales: given a reduced QCD-temperature $t_{\rm
  glue}$, the related Yang-Mills temperature $t_\mrm{YM}$ is that
which minimises \Eq{eq:measure}. This leads to 
\begin{equation}
t_\mrm{YM} (t_\mrm{glue}) \approx 0.57\, t_\mrm{glue}\,,\label{eq:tYMtglue}
\end{equation}
which, together with the absolute temperature scale in
\Eq{eq:Tabsolute} serves as an important input for model
studies. This relation is displayed in Fig.~\ref{fig:tYM_tglue}.
Equation \eq{eq:tYMtglue} also provides the map from the
temperature-dependence of the Yang-Mills order-parameter to that
derived from the glue potential by using $t_\mrm{YM} (t_\mrm{glue}) $
in \Fig{fig:SU3PolFRG}. The respective order-parameter flattens, but
still shows a first-order transition. This procedure also extends to
finite chemical potential e.g.\ with the chemical potential
dependence in Ref.~\cite{Schaefer:2007pw}. A refinement of this estimate
based on full QCD flows, which obeys the Silver-Blaze property for
vanishing temperature, is derived in Ref.~\cite{Herbst:2013ail}.
Note also that the 
approximation \eq{eq:tYMtglue} holds only for small and moderate temperatures, as
at high scales the slope of Eq.~\eqref{eq:tYMtglue} saturates,
where one reaches the perturbative limit and the potentials reach
their asymptotic form. In fact, we would naively expect that the results
for the reduced temperatures agree at (very) high temperatures where the
quark degrees of freedom are parametrically suppressed.

In the following we show how our prescription \eqref{eq:tYMtglue}
can be used to amend PNJL/PQM model studies. More precisely, we use
this prescription to improve the potentials for the Polyakov loop
entering these models. Alternatively to our procedure outlined in the
subsequent sections, we could also directly use the Polyakov-loop
potentials obtained from our functional RG approach in a PNJL/PQM
model study to get even closer to the full QCD study in
Refs.~\cite{Braun:2009gm,Haas:2010bw,Pawlowski:2010ht}. For first studies in
this direction, we refer the reader to
Refs.~\cite{Braun:2011fw,Braun:2012zq} where the ground state of the
nonperturbative Polyakov-loop potential has been used to study the relation
of confinement and chiral symmetry breaking.

\section{\label{sec:PQM}Polyakov-quark-meson model\protect}

\subsection{PQM model with 2+1 quark flavours}

The particle content of the PQM model with 2+1 flavours are the
constituent quarks minimally coupled to gauge fields, and coupled to
mesons via a Yukawa-type term.  The coupling to the mesons generates
the masses of the quarks because the mesonic potential contains
spontaneous and explicit chiral symmetry breaking. The action of the
PQM model with 2+1 flavours includes kinetic terms for the dynamical
low-energy degrees of freedom, quarks and mesons. The quarks are
minimally coupled to a background gauge field $\langle A_0\rangle$
as discussed in Sec.~\ref{sec:FRG} which can be rewritten in terms
of a coupling to the Polyakov loop background, $\Phi[\langle
A_0\rangle]$. The kinetic part of the model is complemented by the
effective potential as a function of the order parameters and
thermodynamic control parameters. In the present work we restrict
ourselves to isospin symmetric matter, and do not distinguish
between the up and down quark sectors. Hence, the order parameters
of chiral symmetry are the light or nonstrange condensate $\sigma$ and
the strange chiral condensate $\sy$ and the pion expectation values
vanish $\vec{\pi}=0$, e.g.\ Refs.~\cite{Lenaghan:2000ey,
  Schaefer:2008hk}. Additionally the pure glue sector in \Eq{eq:QCDflow},
\Fig{fig:QCDflow} leads to an effective potential of the Polyakov loop
$\Phi[\langle A_0\rangle]$. In summary, the effective potential of the model reads 
\begin{multline}
  \Omega\blr{\sigma,\sy,\Phi,\Phib;\,T,\mu_f} = U\blr{\sigma,\sy} +
  \mc{U}\blr{\Phi,\Phib;\,T}+\\ +
  \Omega_\qqb\blr{\sigma,\sy,\Phi,\Phib;\,T, \mu_f} \;, 
	\label{eq:grand_canon_pot}
\end{multline}
where $\bar\Phi$ is the conjugate of $\Phi$. The mesonic
order-parameter potential for chiral symmetry breaking can be
written as follows~\cite{Lenaghan:2000ey, Schaefer:2008hk}
\bea
U\blr{\sx,\sy} &=& \frac{m^2}{2}\blr{\sx^2 +  \sy^{2}} + 
\frac{\lambda_1}{2} \sx^2 \sy^2 + \nonumber\\
&& + \frac{1}{8}\blr{2 \lambda_1 + \lambda_2}\sx^4+
\frac{1}{8}\blr{2 \lambda_1 + 2\lambda_2} \sy^4 -\nonumber\\
&& - \frac{c}{2 \sqrt{2}} \sx^2 \sy -h \sx -h_\mrm{s} \sy \;.
	\label{eq:MesonPot}
\eea 
Hence, the mesonic sector has six parameters that are the coupling
constants $m^2$, $\lambda_1$, $\lambda_2$ and $c$ and the explicit
chiral symmetry breaking terms $h$ and $h_\mrm{s}$. They are adjusted
to the pion and kaon decay constants $f_\pi$ and $f_\mrm{K}$ and the
meson masses of the scalar and pseudoscalar octet, namely $m_\pi$,
$m_\mrm{K}$, $m_\eta^2+m_{\eta'}^2$ and $m_\sigma$. The mass of the
sigma meson is not exactly known. Within our model the sigma meson is
identified with the experimentally measured resonance $f_0(500)$ with
mass $m_{\rm f_0}=\blr{400 - 550}\,\mrm{MeV}$
\cite{Beringer:1900zz}. We will discuss the dependence of our results
on this uncertainty. Once this set of masses and decay constants is
given, the model parameters are defined. For their explicit derivation
we refer to Refs.~\cite{Lenaghan:2000ey, Schaefer:2008hk}. The values
of the constants we use to calculate these parameters are listed in
Table~\ref{tab:chiral_pot_constants}.
\begin{table}
  \caption{Values of constants to which the parameters
    of the mesonic potential are adjusted to, according to 
    Ref.~\cite{Beringer:1900zz} and value of the constituent quark mass 
    of the light (up and down) quarks that we use to fix the 
    quark-meson Yukawa coupling in Eq.~\eqref{eq:quarkmasses_coupling}.}
	\begin{ruledtabular}
		\begin{tabular}{l||ccccccc|c}
                  Constant & $f_\pi$ & $f_\mrm{K}$ & $m_\pi$ & $m_\mrm{K}$ 
                  & $m_\eta$ & $m_{\eta'}$ & $m_\sigma$ & $m_\mrm{l}$ \\ \hline 
                  Value [MeV] & 92 & 110 & 138 & 495 & 548 & 958 & 400-600 & 300 \\
		\end{tabular}
	\end{ruledtabular}
	\label{tab:chiral_pot_constants}
\end{table}

The last term of Eq.~\eqref{eq:grand_canon_pot} represents the
constituent quark sector and derives from the QCD flow equation \eq{eq:QCDflow},
\Fig{fig:QCDflow} within a 1-loop approximation to the quark loop. 
It includes the coupling to the Polyakov-loop variable $\Phi[\langle
A_0\rangle]$ and the mesons
\begin{multline}
	\label{eq:qqbPotential}
	\Omega_\qqb(\sx,\sy,\Phi,\Phib;T,\mu_f)=-2T\sum_{f=\mrm{u,d,s}}\int\frac{d^3p}{(2\pi)^3}\times\\
    	\times\left\{\ln\left[1 + 3\blr{\Phi + \Phib e^{-\blr{E_f-\mu_f}/T}}\times  \right.\right. \\
    	\left.\times e^{-\blr{E_f-\mu_f}/T} + e^{-3\blr{E_f-\mu_f}/T}\right] + \\
	+ \ln\left[1 + 3\blr{\Phib + \Phi e^{-\blr{E_f+\mu_f}/T}}\times \right. \\
	\left.\left.\times e^{-\blr{E_f-\mu_f}/T} +
            e^{-3\blr{E_f+\mu_f}/T}\right]\right\}\;.\qquad
\end{multline}
The dispersion relation of the quarks,
\be E_{f} = \sqrt{k^2+m_f^2}\;,
\ee 
couples the chiral condensates to the Polyakov-loop variables since
the constituent light and strange quark masses are, respectively,
\be
	\label{eq:quarkmasses_coupling}
	m_\mrm{l} = \frac{g}{2} \sx \qquad \text{and} \qquad m_\mrm{s}
        = \frac{g}{\sqrt{2}} \sy\;.
\ee
The Yukawa coupling, $g$, in \Eq{eq:quarkmasses_coupling} between
quarks and (pseudo-)scalar mesons is fixed by choosing the constituent
mass of the light quarks to be $m_\mrm{l} = 300\,\mrm{MeV}$, which
results in $m_\mrm{s} \simeq 417\,\mrm{MeV}$ for the constituent
strange quarks.

One can combine nonstrange or light and strange condensate to what is
called the subtracted condensate
\be
	\label{eq:subcond}
	\Delta_\mrm{l,s} = \frac{\left.\sx-\frac{h}{h_\mrm{s}}\sy\
          \right|_T}{\left.\sx-\frac{h}{h_\mrm{s}}\sy\ \right|_{T=0}}
        \:.  
\ee
This quantity is better accessible in lattice simulations than the
condensates themselves and hence will be used for the comparison of
the behaviour of the chiral sector in the PQM model with lattice
calculations.

It can be easily inferred from the FRG-setting discussed in
Sec.~\ref{sec:FRG} that the effective potential detailed above has
to change under a variation of the physical UV-scale $\Lambda$ of
the low-energy model. This follows from the fact that $\Lambda$ also plays the r$\hat{\rm
o}$le of an IR scale, up to which QCD fluctuations are integrated
out, see Sec.~\ref{sec:FRG}. In an FRG-setting the
$\Lambda$-dependence can be directly computed from the flow and
follows from the invariance of the effective action at vanishing
cutoff scale under a variation of $\Lambda$, see the reviews
\cite{Litim:1998nf, Berges:2000ew, Polonyi:2001se, Pawlowski:2005xe,
  Gies:2006wv, Schaefer:2006sr,Pawlowski:2010ht,Schaefer:2011pn,
  Braun:2011pp,vonSmekal:2012vx}. In the model setup this
well-known fact is related to the so-called vacuum (or sea)
contribution to the effective potential, see
Ref.~\cite{Skokov:2010sf}. Its origin in the FRG-setting makes clear that
it already includes part of the quantum fluctuations related to the
UV-scale $\Lambda$. Consequently we expect significant contributions
in particular for larger chemical potential, large mass scales and
higher temperature. The impact on the effective potential in the PQM
model is discussed in more detail in Ref.~\cite{Herbst:2013ail}. In the present work
it is a subleading effect as we only consider vanishing densities
and adopt low values for the mass of the $\sigma$-meson.

\subsection{\label{ssec:PloopPot}Polyakov-loop potential\protect}

The Polyakov-loop potential $\mc{U}\blr{\Phi,\Phib;T}$ simply
is the pure glue potential $V_{\rm glue}[\langle A_0\rangle]$
defined in \Eq{eq:Vglue}. However, in the standard approach to
Polyakov-loop extended models this potential is usually modelled in the
following way: The functional form for the potential contains all
terms invariant under $Z_3$ transformations up to quartic order in the
Polyakov loop (see e.g.~Refs.~\cite{Pisarski:2000eq, Scavenius:2002ru,
  Fukushima:2003fw, Ratti:2005jh, Roessner:2006xn}). One possible
parametrisation of the Polyakov-loop potential is the polynomial
parametrisation of Refs.~\cite{Scavenius:2002ru,
  Ratti:2005jh}. Reference~\cite{Scavenius:2002ru} used a different
definition of the coefficients than Ref.~\cite{Ratti:2005jh}. A simple
calculation allows the translation from one set of coefficients to the
other, see Ref.~\cite{Mintz:2012mz}.
\begin{multline}
  \frac{\mc{U}_{\mrm{poly}}\blr{\Phi,\Phib,t}}{T^{4}} =
  -\frac{b_2\blr{t}}{2} \Phib\Phi - \frac{b_3}{6}
  \blr{\Phi^3+\Phib^3}+\\ +\frac{b_4}{4} \blr{\Phib\Phi}^2\;,
        \label{eq:Polyakov_potential_polynomial}
\end{multline}
with the temperature-dependent coefficient $b_2$ defined as
\bea b_2\blr{t} &=& a_0 + \frac{a_1}{1+t} + \frac{a_2}{\blr{1+t}^2} +
\frac{a_3}{\blr{1+t}^3}\;.  \eea
Another possible parametrisation of the Polyakov-loop potential of
Ref.~\cite{Roessner:2006xn} contains a logarithmic term that restricts the
Polyakov-loop expectation value to be smaller than one,
\pagebreak
\begin{multline}
	\label{eq:PloopPot}
	\frac{\mc{U}_\mrm{log}\blr{\Phi,\Phib,t}}{T^4} =
        -\frac{A \blr{t}}{2} \Phi\Phib + B\blr{t} \times  \\
        \times \ln \left[ 1 - 6 \Phi\Phib + 4 \blr{\Phi^{3}+\Phib^{3}}
          - 3 \blr{ \Phi\Phib }^{2}\right] \;,
\end{multline}
where both coefficients are temperature dependent, 
\bea
A\blr{t} &=& A_0 + \frac{A_1}{1+t} + \frac{A_2}{\blr{1+t}^2} \;,\\
B\blr{t} &=& \frac{B_3}{\blr{1+t}^3} \;.
\eea
The parameters of the model potential are then determined with the
help of Yang-Mills lattice data \cite{Boyd:1996bx,Kaczmarek:2002mc} for the Polyakov-loop
expectation value $\langle \Phi\rangle$, that is the location of the
absolute minimum, and the thermodynamics, that is the value of the
potential at the minimum. They are listed in
Table~\ref{tab:Ploop_pot_params}.
\begin{table}
	\caption{Parameters of the gauge potential parametrisations
          for fits to the lattice Yang-Mills simulations \cite{Boyd:1996bx,Kaczmarek:2002mc}.}
	\begin{ruledtabular}
		\begin{tabular}{lcccccc}
			        						& $a_0$	& $a_1$	& $a_2$	& $a_3$	& $b_3$	& $b_4$	\\
			Poly-I \cite{Scavenius:2002ru} 	& 1.53 	& 0.96  	& -2.3	& -2.85	& 13.34	& 14.88 	\\
			Poly-II \cite{Ratti:2005jh}      		& 6.75  	& -1.95 	&2.625 	& -7.44 	& 0.75   	 & 7.5 \\
			\hline
			 						& $A_0$	& $A_1$	& $A_2$	& $B_3$ \\
			Log \cite{Roessner:2006xn} 		& 3.51 	& -2.47 	& 15.2 	&-1.75
		\end{tabular}
	\end{ruledtabular}
	\label{tab:Ploop_pot_params}
\end{table}

\begin{figure}[b]
	\centering
	\includegraphics[width=0.47\textwidth,]{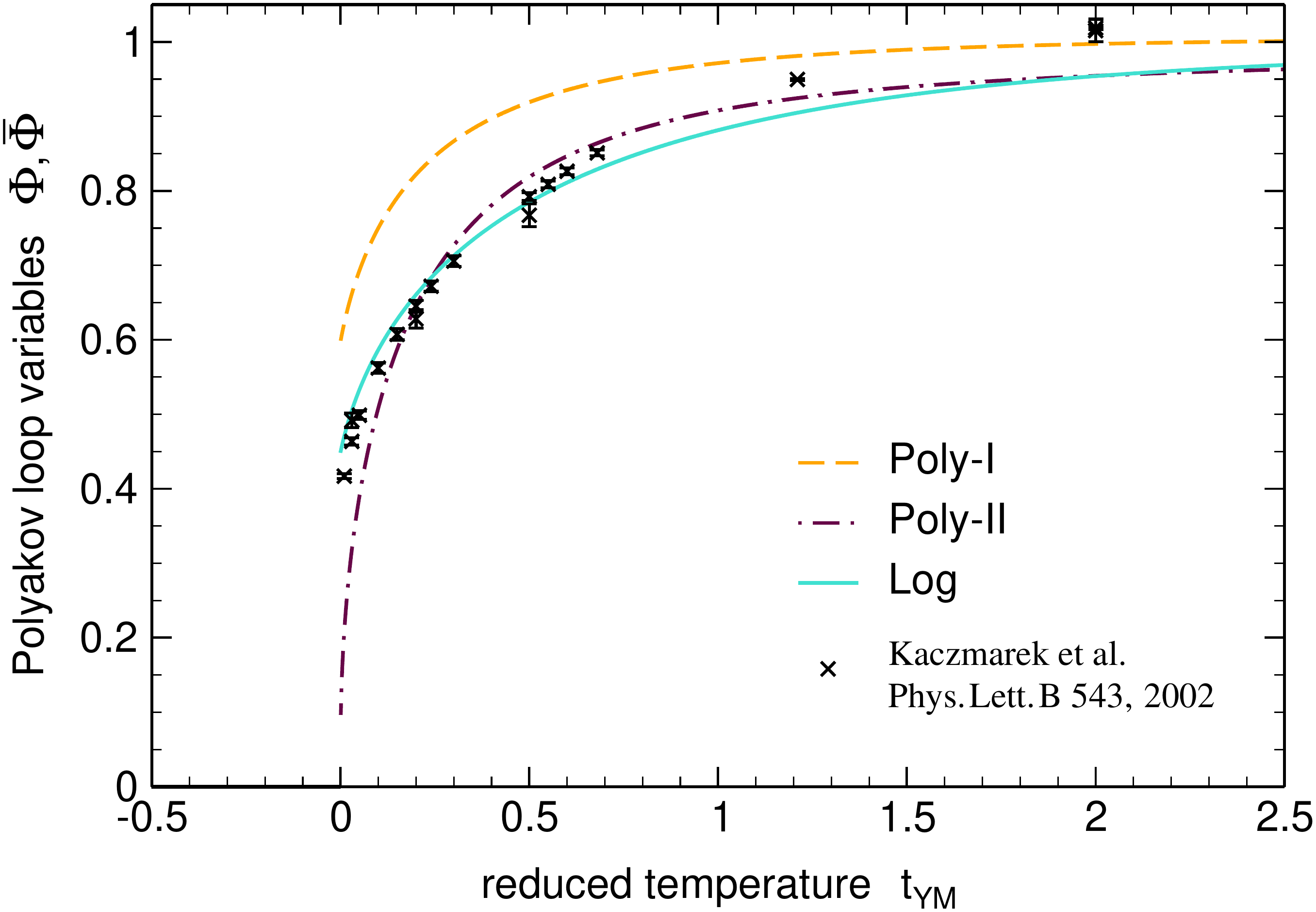}
	\caption[]{Polyakov loop as a function of the reduced
          temperature in the pure-gauge sector for the different
          parametrisations and parameter sets of Table
          \ref{tab:Ploop_pot_params}, compared to corresponding
          lattice results taken from Ref.~\cite{Kaczmarek:2002mc}.}
	\label{fig:PloopYM}
\end{figure}

\begin{figure}
	\centering
	\includegraphics[width=0.47\textwidth,]{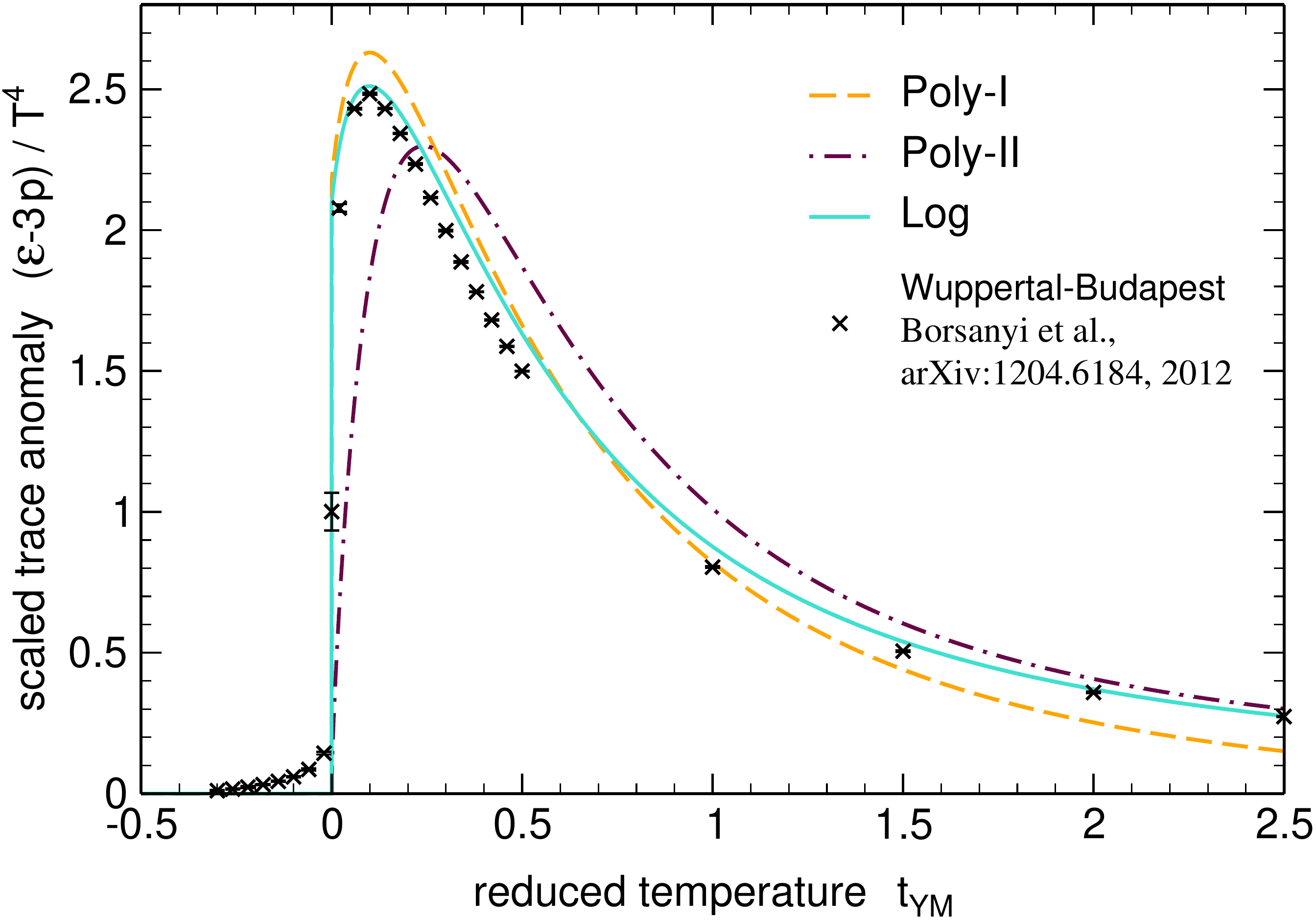}
	\caption[]{Scaled trace anomaly as a function of the reduced
          temperature in the pure-gauge sector for the different
          parametrisations and parameter sets of Table
          \ref{tab:Ploop_pot_params}, compared to corresponding
          lattice results taken from Ref.~\cite{Borsanyi:2012ve}.}
	\label{fig:e3pYM}
\end{figure}

To compare the different parametrisations and parameter sets and for
the later discussion of their impact we show in
Figs.~\ref{fig:PloopYM} and \ref{fig:e3pYM} the location of the minima
of the potential and the normalised trace anomaly.
We see that, first,
the logarithmic parametrisation is in best agreement with the lattice
data, second, the Poly-I potential features the strongest phase
transition, third, the Poly-II potential leads to a relatively weak
first-order transition and shows an offset of the peak of the trace
anomaly compared to the other two potentials and the lattice data.
The different description of the strength of the phase transition
shows a significant impact on the surface tension in the PQM model at finite density \cite{Mintz:2012mz,Mintz:2013}.

The construction of the Polyakov-loop potential $\CU$ entails that it
models the pure gauge potential
$\mc{U}_\mrm{YM}\blr{t_\mrm{YM}}/T_\mrm{YM}^4$ in terms of the
variable $\langle \Phi\rangle$. It describes successfully the first
order phase transition of SU(3) Yang-Mills theory. The critical temperature of
the Polyakov-loop potential entering the reduced temperature is
accordingly that of pure gauge theory,
$T_\mrm{cr}^\mrm{YM}=270\,\mrm{MeV}$.

We conclude that the use of the model potentials $\CU$ incorporates
several approximations and deficiencies: Firstly, a Yang-Mills
potential is used instead of the glue potential. Secondly, it only
models the Polyakov-loop potential in Yang-Mills theory by using the
location of its absolute minimum and the value at the minimum. Below
$T_c$ this basically provides no information about the
potential. Thirdly, the potential has to be known as a function of
$\Phi[\langle A_0\rangle]$ as the coupling to the matter sector is
described in this variable. Strictly speaking, $\Phi[\langle
A_0\rangle]\approx \langle \Phi\rangle$ is only valid in the Gau\ss
ian approximation, in general we have $\Phi[\langle A_0\rangle] \geq
\langle \Phi\rangle$ up to renormalisation issues, see
Refs.~\cite{Braun:2007bx,Marhauser:2008fz}. 

In the present work we resolve the first approximation by using the
relation between pure gauge and glue effective potential
\eqref{eq:tYMtglue}. This allows us to convert the Yang-Mills
potential $\CU_{\rm YM}$ for the Polyakov loop to a glue potential of
full QCD,
\bea
\label{eq:gluePotYMPot}
\frac{\mc{U_\mrm{glue}}}{T^4} \blr{\Phi,\Phib,t_\mrm{glue}}&=&
\frac{\mc{U_\mrm{YM}}}{T_\mrm{YM}^4}\blr{\Phi,\Phib,t_\mrm{YM}(t_{\rm
    glue})} \;, \eea
%
with $t_\mrm{YM}(t_{\rm glue})$ in \Eq{eq:tYMtglue}. The absolute
temperature scale is then determined by the value of the transition
temperature \eq{eq:Tabsolute}. The resolution of the other two
approximations will be discussed elsewhere.

\section{\label{sec:ResultsDis}Results and Discussion\protect}

Given the temperature $T$ and the quark chemical potentials $\mu_f$,
the effective potential \eqref{eq:grand_canon_pot} is given as a
function of the four order-parameters $\sigma$, $\sigma_\mrm{s}$,
$\Phi$ and $\Phib$. In equilibrium, the expectation values of the
order parameters are given by extremising the effective potential 
\bea
	\label{eq:gap_eqs}
	\frac{\pd\Omega}{\pd\sx} = \frac{\pd\Omega}{\pd\sy} =
        \frac{\pd\Omega}{\pd\Phi} = \frac{\pd\Omega}{\pd\Phib} &=& 0
        \;.  
\eea 
For vanishing density the EoMs in \Eq{eq:gap_eqs}, with the constraint
$\Phi=\bar\Phi$, are given by a minimum of the effective potential. In
contradistinction, for nonvanishing density, with $\Phi\neq \bar\Phi$
the solution of Eqs.~\eqref{eq:gap_eqs} is only a saddle point, see
the discussions in Refs.~\cite{Fukushima:2006uv,Schaefer:2007pw,Mintz:2012mz}.

In this work we compare the solutions of the equations of motion
\eqref{eq:gap_eqs} using for the Polyakov-loop potential the pure YM
potential and the enhancement to a glue potential
\eqref{eq:gluePotYMPot} by the relation \eqref{eq:tYMtglue}.

The lattice results we compare our results to are the continuum
extrapolations of Refs.~\cite{Borsanyi:2010bp,Borsanyi:2010cj} and those
of the HotQCD collaboration using the HISQ action and temporal lattice
extent $N_\tau = 12$ (dots) and $N_\tau = 8$ (squares) of
Refs.~\cite{Bazavov:2011nk, Bazavov:2012bp}, both with physical quark
masses.

To investigate the improvement of the Polyakov-loop potential by
Eqs.~\eqref{eq:gluePotYMPot} and \eqref{eq:tYMtglue} we first show
results with the combination of glue critical temperature, mass of the
sigma meson and parametrisation of the Polyakov-loop potential that
reproduces best the lattice results. Afterwards, we discuss the
dependence of our results on the uncertainties of these parameters.

\subsection{\label{ssec:Res_ImPoLoPot}Improved Polyakov-loop
  potential\protect}

To compare the results of the PQM model using the Yang-Mills
Polyakov-loop potential and the quark-improved Polyakov-loop potential
we first have to choose a canonical setting for the free parameters
and uncertainties that are the critical temperature and the
parametrisation of the Polyakov-loop potential as well as the mass of
the $\sigma$-meson.

\begin{figure}
	\centering
	\includegraphics[width=0.47\textwidth,]{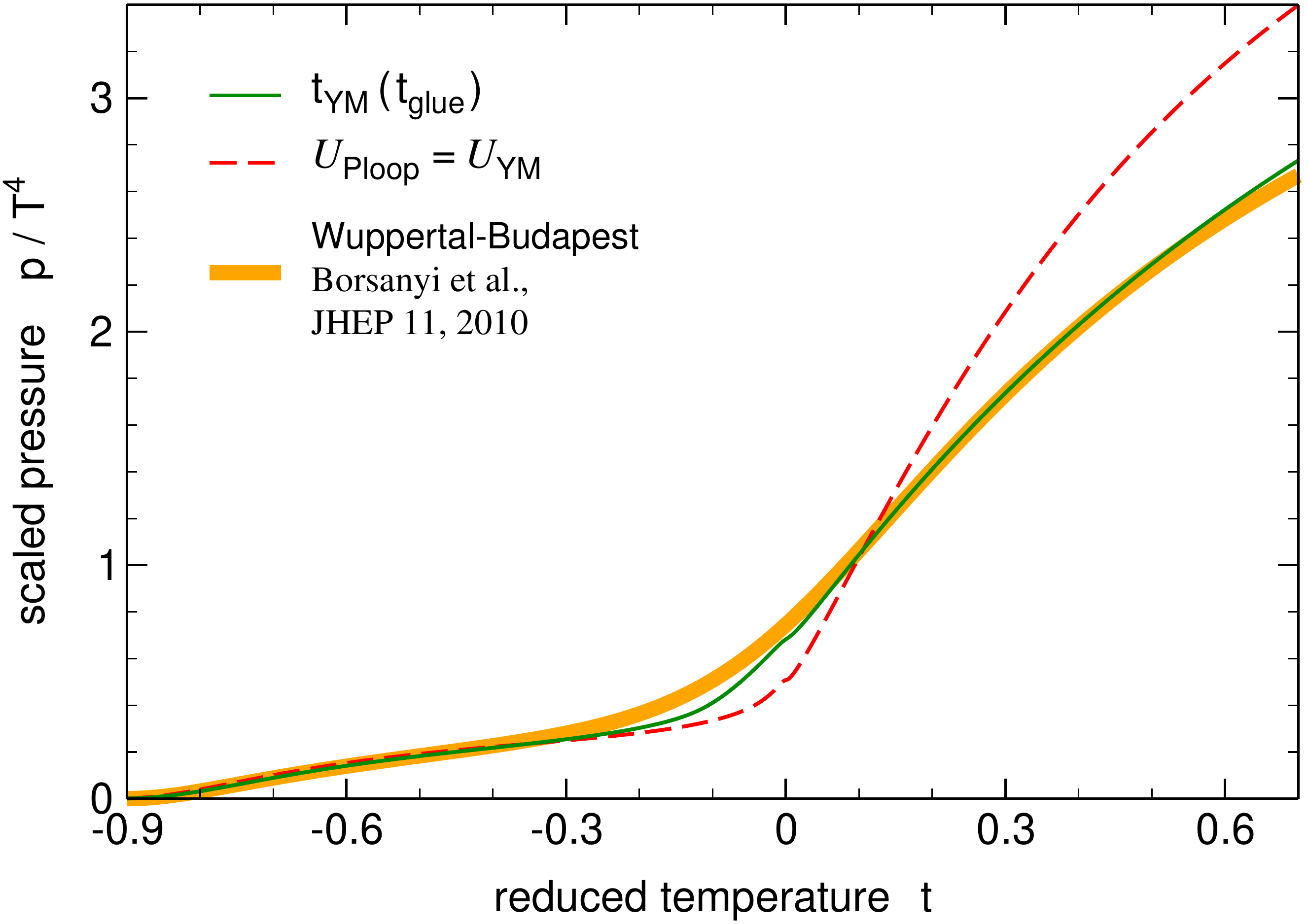}
	\caption[]{Scaled pressure $p/T^4$ as a function of
          temperature at $\mu_f=0$. The red dashed line is the result
          using as Polyakov-loop potential the pure Yang-Mills
          potential, the full green line is the case for the
          improvement to the glue potential. The orange band is the
          fit to the continuum extrapolation of the lattice results of
          Ref.~\cite{Borsanyi:2010cj}.}
	\label{fig:Pressure}
\end{figure}

The logarithmic parametrisation of the
Polyakov-loop potential \eqref{eq:PloopPot} with the parameters of
Ref.~\cite{Roessner:2006xn} is in best agreement with the expectation
value of the Polyakov loop and the equation of state of Yang-Mills theory as
can be seen in Figs.~\ref{fig:PloopYM} and \ref{fig:e3pYM}. Therefore,
we use this parametrisation for the following discussion. Furthermore,
we take a medium value for the mass of the sigma meson of
$m_\sigma=500\,\mrm{MeV}$.  To adjust the critical temperature of the
glue effective potential in full QCD we want to bring our model in
agreement with lattice results of the normalised pressure $p/T^4$
since the pressure is directly related to basic quantities in lattice
simulations (the free energy) and model studies (the minimum of the
potential).  Figure \ref{fig:Pressure} shows that with
$T^{\text{glue}}_{\text{cr}}=210$\,MeV we find a normalised pressure
$p/T^4$ that is in very good agreement with the continuum extrapolated
lattice results of Ref.~\cite{Borsanyi:2010cj}.

Now that we have fixed all uncertainties we can investigate how the
results evolve when we amend the pure Yang-Mills Polyakov-loop potential to a
glue potential \eqref{eq:gluePotYMPot} by the relation
\eqref{eq:tYMtglue}.  In Fig.~\ref{fig:Pressure} we see that the
enhancement of the Polyakov-loop potential smoothes the phase
transition significantly bringing the effective model in close
agreement with lattice and functional calculations.  This agreement
holds in the low as well as in the high temperature phase. To achieve
this agreement in the phase were chiral symmetry is broken we took
into account the contribution of thermal pions.

The critical temperature of the glue potential
$T^{\text{glue}}_{\text{cr}}=210$\,MeV for which we find the best
agreement with the lattice calculation is larger than the estimate for
2+1 quark flavours of Ref.~\cite{Schaefer:2007pw}.  But this estimate
of $T^{\text{glue}}_{\text{cr}}\blr{2+1}=182$\,MeV is only a lower
limit as outlined in Sec.~\ref{sec:FRG} when discussing Eq.~\eq{eq:Tabsolute}. We find comparable curves for the combinations
($m_\sigma=400\,\mrm{MeV}$,
$T^{\text{glue}}_{\text{cr}}=180\,\mrm{MeV}$) and
($m_\sigma=600\,\mrm{MeV}$,
$T^{\text{glue}}_{\text{cr}}=250\,\mrm{MeV}$). We will discuss the
impact of the critical temperature of the Polyakov-loop potential and
of the mass of the $\sigma$-meson in detail in the following sections.

\begin{figure}
	\centering
	\includegraphics[width=0.47\textwidth,]{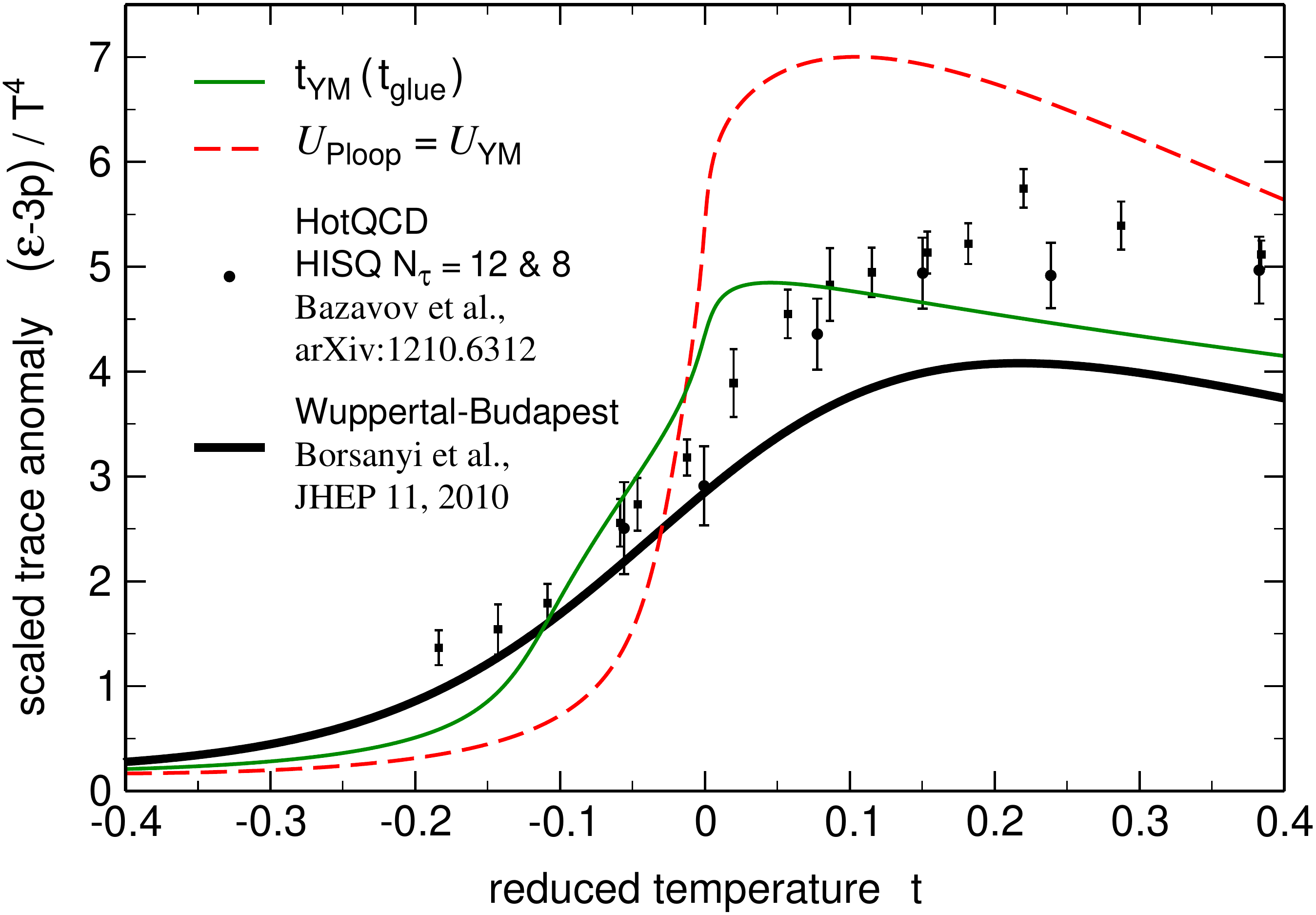}
	\caption[]{Scaled trace anomaly $\blr{\epsilon-3p}/T^4$ as
          function of temperature at $\mu_f=0$. The red dashed line is
          the result using as Polyakov-loop potential the pure
          Yang-Mills potential, the full green line when the
          improvement to the glue potential by the adjustment from
          functional calculations is taken into account. The black
          band is the fit to the continuum extrapolation of the
          lattice results of Ref.~\cite{Borsanyi:2010cj} and the data
          points lattice results of the HotQCD collaboration
          \cite{Bazavov:2012bp}.}
	\label{fig:TraceAnomaly}
\end{figure}

To complete the discussion of the pressure shown in
Fig.~\ref{fig:Pressure} let us note that the result of the PQM model
with the quark-improved Polyakov-loop potential still shows a
transition that is marginally steeper than that on the lattice.  A
quantitative probe of this slight deviation is the trace anomaly or
interaction measure which tests not only the free energy or minimum of
the potential but contains as well information about their temperature
derivative. In Fig.~\ref{fig:TraceAnomaly} we compare results for the
normalised trace anomaly $\blr{\epsilon-3p}/T^4$. Even though the
agreement of the normalised pressure of the effective model using the
improved glue potential and the lattice calculation is remarkably
close the slight differences get more visible in the trace anomaly.
The amplitude is in good agreement with lattice results, nevertheless
the transition in the effective model is still steeper than in lattice
calculations. However, the calculations done here within a basic
effective model neglect contributions from correlations. Taking those
into account smoothes the phase transition, as is shown e.g.\ in
Refs.~\cite{Herbst:2010rf, Schaefer:2011ex} which will lead to a better
agreement.

Anyhow, one sees in the interaction measure that the improved glue
potential smoothes the transition compared to the standard calculation
with the Yang-Mills Polyakov-loop potential. So the emergence of the quark
degrees of freedom happens in a larger temperature interval and its
amplitude is in better agreement with lattice results.

After this analysis of thermodynamic quantities, let us now analyse
the evolution of the order parameters that are the Polyakov loop for
deconfinement and the subtracted chiral condensate (\ref{eq:subcond})
for chiral symmetry restoration.

\begin{figure}
	\centering
	\includegraphics[width=0.47\textwidth]{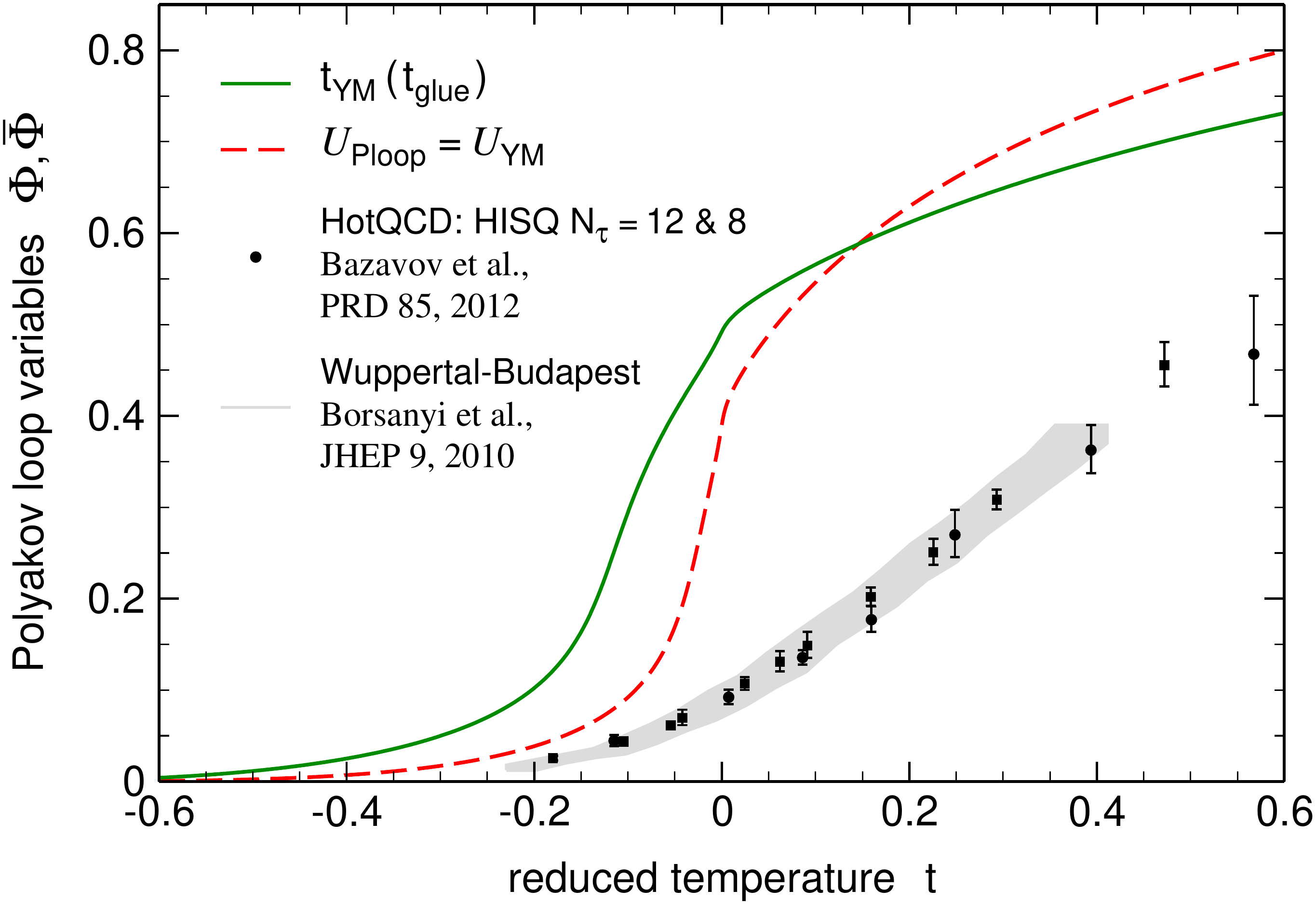}
	\caption[]{The Polyakov-loop expectation value $\Phi$ as
          function of temperature at $\mu_f=0$. The red dashed line is
          the result using as the Polyakov-loop potential the pure
          Yang-Mills potential, the full green line when the
          improvement to the glue potential by the adjustment from
          functional calculations is taken into account. The grey band
          is the continuum extrapolation of the lattice results of
          Ref.~\cite{Borsanyi:2010bp} and the data points lattice
          results of Ref.~\cite{Bazavov:2011nk}.}
	\label{fig:Ploop}
\end{figure}

The biggest discrepancy between Polyakov-loop extended effective
models and lattice results for full QCD is seen in the Polyakov-loop
expectation value \cite{Schaefer:2009ui}. The lattice data shows a
smoother transition with significant smaller values. This can be seen
in Fig.~\ref{fig:Ploop} and holds as well when the quark-improved
Polyakov-loop potential is applied. Including contributions of
fluctuations to the model will certainly reduce this large offset to
the lattice data. 
However, a part of the discrepancy originates in the inherent
  approximations which are still present: the derivation of the PQM
  model from QCD entails that the Polyakov-loop variable in the quark
  loop is $\Phi\left[\langle{A_0}\rangle\right]$ and not
  $\langle\Phi\rangle$ as used in the Polyakov-loop--model potentials
  $\CU$. This mismatch can be resolved by using the QCD glue potential
  $V_{\rm glue}$ as defined in \Eq{eq:Vglue}. Then the model is fully
  consistent with its QCD counterpart defined by the flow equation
  \eq{eq:QCDflow}. The Polyakov loop observable accessible in such a
  continuum approach to full QCD is
  $\Phi\left[\langle{A_0}\rangle\right]$ and not $\langle\Phi\rangle$
  computed on the lattice. The former observable has been computed in
  Ref.~\cite{Braun:2009gm} within QCD-flows for two-flavour QCD in the
  chiral limit. It shows a rather smooth transition but cannot be
  directly compared with present lattice data. On the lattice, the
  chiral limit is not yet accessible. In any case, the two observables
  are different and mapping them into each other is an interesting open
  issue. 

One sees in Fig~\ref{fig:Ploop} that by applying the improved glue
potential, the Polyakov loop is shifted differently in the two phases,
as anticipated in Figs.~\ref{fig:GlueYMPotentials} and
\ref{fig:tYM_tglue} and Eq.~(\ref{eq:tYMtglue}). So, the Polyakov-loop
variables are shifted to higher expectation values in the confined
phase and to lower values in the deconfined phase.  This shift in the
respective phases is what leads to a overall smoother transition, not
only of the Polyakov-loop expectation value.  The evolution of the
Polyakov loop when the quark-improved potential is considered has the
advantage that the Polyakov-loop variables are shifted to a value of
$\sim0.5$ at the transition which is a reasonable criterion to define
the transition between the confined phase and the quark gluon plasma
in case of a crossover.

\begin{figure}
	\centering
	\includegraphics[width=0.47\textwidth]{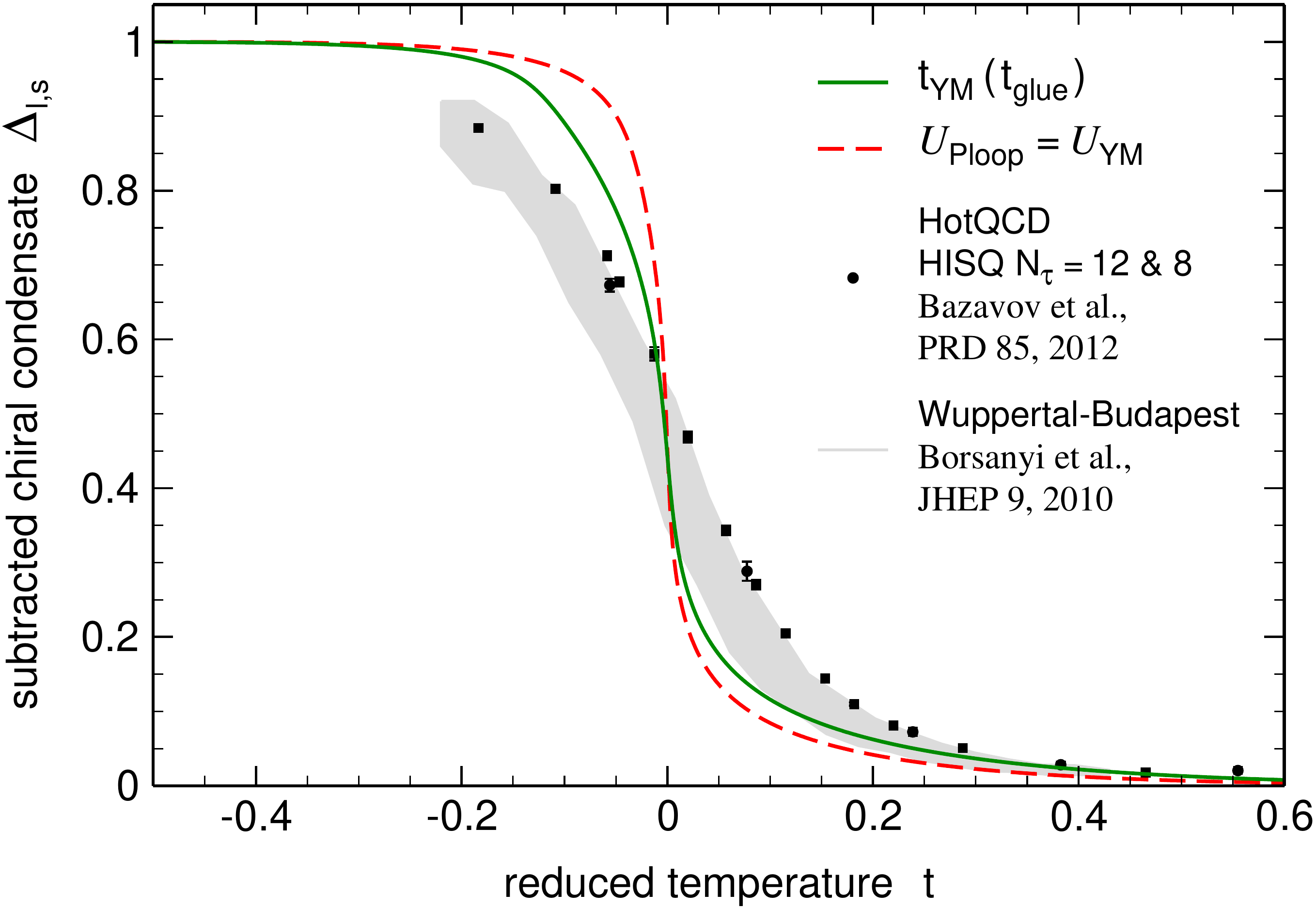}
	\caption[]{The subtracted chiral condensate $\Delta_\mrm{l,s}$
          as a function of temperature at $\mu_f=0$. The red dashed
          line is the result using as Polyakov-loop potential the pure
          Yang-Mills potential, the full green line when the
          improvement to the glue potential by the adjustment from
          functional calculations is adopted. The grey band is the
          continuum extrapolation of the lattice results of
          Ref.~\cite{Borsanyi:2010bp} and the data points lattice
          results of Ref.~\cite{Bazavov:2011nk}, see also Ref.~\cite{Fischer:2012vc}.}
	\label{fig:subcond}
\end{figure}

The adjustment of the Polyakov-loop potential from the pure gauge
potential to the glue potential in full QCD does not only affect the
Polyakov-loop expectation value but also quarks and mesonic degrees of
freedom because of the nontrivial coupling to the gauge field.
Therefore, we show in Fig.~\ref{fig:subcond} how the subtracted chiral
condensate evolves with increasing temperature for vanishing
density. The result of the PQM model using as Polyakov-loop potential
the pure gauge potential and the improvement to the glue potential
from functional calculations show both a steeper decrease in the
transition region than the lattice calculations. Nevertheless, the
adjustment of the gauge potential to the glue potential leads to a
smoother chiral transition and to an improvement towards the result of
lattice
calculations.

In the previous and following figures the absciss{\ae} are in units of
the reduced temperature of full QCD $t = (T-T_\mrm{c})/T_\mrm{c}$. To
have a unique criterion for lattice and model calculations we determine
the pseudocritical temperature $T_\mrm{c}$ of the crossover transition
by the peak of the susceptibility of the subtracted chiral condensate.
Table \ref{tab:critTemps} summarises the hereby defined pseudocritical
temperatures for the lattice calculations and the model calculations
with the Yang-Mills and improved glue Polyakov-loop potential.
\begin{table}
  \caption{Pseudocritical temperatures for the crossover 
    phase transition at $\mu_f=0$. They are determined by the 
    peaks in the temperature derivatives of the subtracted condensate $\Delta_\mrm{l,s}$.}
	\begin{ruledtabular}
		\begin{tabular}{l|c|c|c|c}
						
						& $\mc{U}_\mrm{YM}$	& $\mc{U}_\mrm{glue}$	& lattice Wuppertal-				& lattice HotQCD \\
						&					&					& Budapest \cite{Borsanyi:2010bp} 	& $N_\tau=12$ \& $8$ \cite{Bazavov:2011nk} \\ \hline
			$T_\mrm{c}$ [MeV]	& 168				& 158				& 157						& 159 \& 163
		\end{tabular}
	\end{ruledtabular}
	\label{tab:critTemps}
\end{table}
We see that applying the quark-improved Polyakov-loop potential leads
to a reduction of the pseudocritical temperature. With
$T_\mrm{c}=158\,\mrm{MeV}$ we get a result that is in very good
agreement with the best lattice calculations. That is another
important ingredient to find the nice agreement of our results for the
equation of state and order parameters.

We conclude that adjusting the gauge potential to the glue
potential in full QCD by applying Eqs.~\eqref{eq:gluePotYMPot} and
\eqref{eq:tYMtglue} improves the description of the chiral and
(de)confinement phase transition with effective models.

\subsection{\label{ssec:ResT0dep}Dependence on the glue critical
  temperature\protect}

In the last section we have showed that replacing the pure Yang-Mills
Polyakov-loop potential by the quark-improved potential from
functional calculations smoothes the phase transition significantly
and brings it to closer agreement with lattice and functional
calculations.  Now, we will investigate the dependence of our results
on the critical temperature of the glue potential
$T^{\mrm{glue}}_\mrm{cr}$. It is not yet uniquely defined so that we
consider it as a parameter. According to the discussion in
Sec.~\ref{sec:FRG} around Eq.~\eq{eq:Tabsolute} we
expect it to be in the interval $\blr{180 - 270}\,\mrm{MeV}$. For this
investigation we use again the logarithmic parametrisation of the
Polyakov-loop potential and a medium sigma meson mass of
$m_\sigma=500\,\mrm{MeV}$.  For purpose of illustration we omit in the
following figures to show the results with the Yang-Mills Polyakov-loop
potential but choose the same range of the ordinates as in the
previous section to give an estimate of the reduction due to the
application of the quark-improved Polyakov-loop potential. We show the
results with both potentials in the same plots elsewhere
\cite{Stiele:2013gra}.

\begin{figure}
	\centering
	\includegraphics[width=0.47\textwidth]{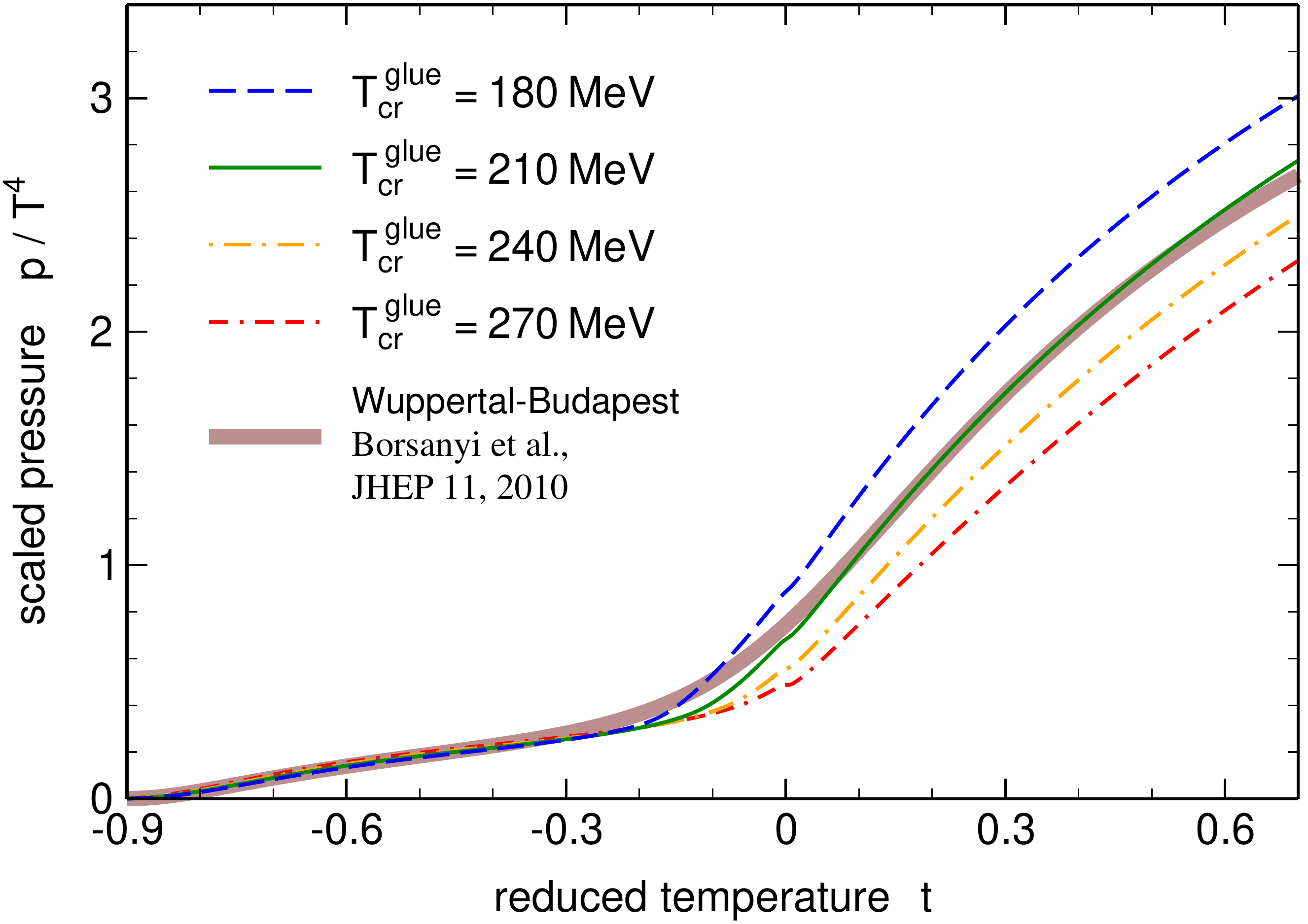}
	\caption[]{Scaled pressure $p/T^4$ as a function of
          temperature at $\mu_f=0$ with the adjustment to the glue
          potential for four different glue critical temperatures. The
          brown band is the fit to the continuum extrapolation of the
          lattice results of Ref.~\cite{Borsanyi:2010cj}.}
	\label{fig:Pressure_T0s}
\end{figure}
Let us start again with the discussion of the normalised pressure
$p/T^4$. We see in Fig.~\ref{fig:Pressure_T0s} that the pressure rises
already at smaller relative temperatures and that it shows a slightly
steeper increase when the critical temperature of the Polyakov-loop
potential is lowered.
Together with the respective pseudocritical temperatures that are
summarised in Table \ref{tab:critTemps_T0s} the best agreement with
lattice results shows the curve with a critical temperature of the
glue potential of $T^{\text{glue}}_{\text{cr}}=210$\,MeV.  As
discussed in the previous section the pseudocritical temperature of
$T_\mrm{c}=158\,\mrm{MeV}$ for
$T^{\mrm{glue}}_\mrm{cr}=210\,\mrm{MeV}$ is in accordance with the
lattice result. But also the value for
$T^{\mrm{glue}}_\mrm{cr}=180\,\mrm{MeV}$ of $T_\mrm{c}=152\,\mrm{MeV}$
is on the lower end of the transition region of lattice calculations
\cite{Borsanyi:2010bp}.  As addressed already in the previous section
including fluctuations to the model smoothes the phase transition
additionally \cite{Herbst:2010rf, Schaefer:2011ex} and slightly increases the
transition temperature as shows e.g.\ a comparison of Refs.\
\cite{Schaefer:2007pw} and \cite{Herbst:2010rf}. So we can expect that taking
into account fluctuations of quarks and mesons would decrease the
transition temperature of the glue potential for which we find best
agreement with the lattice calculation towards the 2+1 flavour estimate
of Ref.~\cite{Schaefer:2007pw}. We will address the quantitative analysis
in a future work.

\begin{figure}
	\centering
	\includegraphics[width=0.47\textwidth]{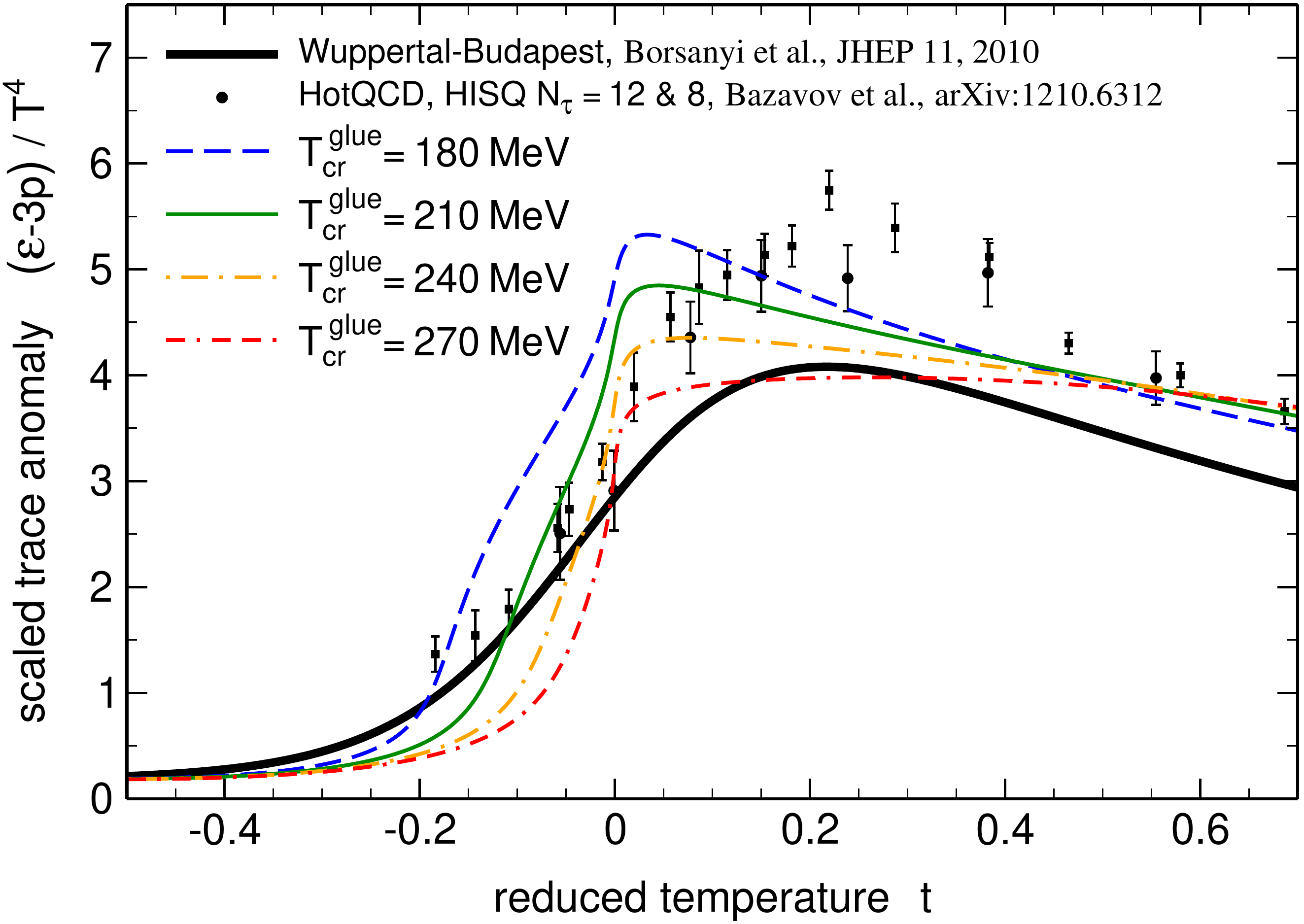}
	\caption[]{Scaled trace anomaly $\blr{\epsilon-3p}/T^4$ as
          function of temperature at $\mu_f=0$ with the adjustment to
          the glue potential for four different glue critical
          temperatures. The black band is the fit to the continuum
          extrapolation of the lattice results of
          Ref.~\cite{Borsanyi:2010cj} and the data points lattice
          results of the HotQCD collaboration~\cite{Bazavov:2012bp}.}
	\label{fig:TraceAnomaly_T0s}
\end{figure}

For the detailed investigation of the steepness of the transition in
dependence of the critical temperature of the improved glue potential
we show in Fig.~\ref{fig:TraceAnomaly_T0s} the normalised trace
anomaly or interaction measure.
\begin{table}
  \caption{Pseudocritical temperatures for the crossover transition at $\mu_f=0$ for different critical temperatures of the glue potential. They are determined by the peaks of  the chiral susceptibility $\pd{\Delta_\mrm{l,s}}/\pd{T}$.}
	\begin{ruledtabular}
		\begin{tabular}{l|c|c|c|c}
			$T^{\mrm{glue}}_\mrm{cr}$ [MeV]	& 180 & 210 & 240 & 270 \\ \hline
			$T_\mrm{c}$ [MeV] 					& 152 & 158 & 164 & 171 \\
		\end{tabular}
	\end{ruledtabular}
	\label{tab:critTemps_T0s}
\end{table}
We see that using our adjustment of the Polyakov-loop potential
towards the glue potential of full QCD the overall amplitude is in
good agreement with lattice calculations.  Like anticipated in the
discussion of the pressure the steepness of the transition increases
when the critical temperature of the Polyakov-loop potential is
lowered. Furthermore, the transition region broadens towards lower
relative temperatures with decreasing transition temperature of the
gauge potential.

An important aspect that we see in Fig.~\ref{fig:TraceAnomaly_T0s} is
that the slope of the trace anomaly becomes smaller and unnatural
above $T_\mrm{c}$ for large critical temperatures of the Polyakov-loop
potential. This behaviour has two reasons.  First, the pseudocritical
temperature is shifted towards the lower end of the transition region
for larger transition temperatures of the glue potential, so that the
transition region extends well above the pseudocritical
temperature. So the emergence of the quark degrees of freedom holds on
above $T_\mrm{c}$ and avoids a decrease of the interaction measure.
Second, the saturation scale of our adjustment between pure YM
Polyakov-loop potential and quark-improved glue potential, i.e., the
upper limit of validity of Eq.~(\ref{eq:tYMtglue}), is closer to the
transition scale for larger critical temperatures of the Polyakov-loop
potential, see Table \ref{tab:critTemps_T0s}.

\begin{figure}
	\centering
	\includegraphics[width=0.47\textwidth]{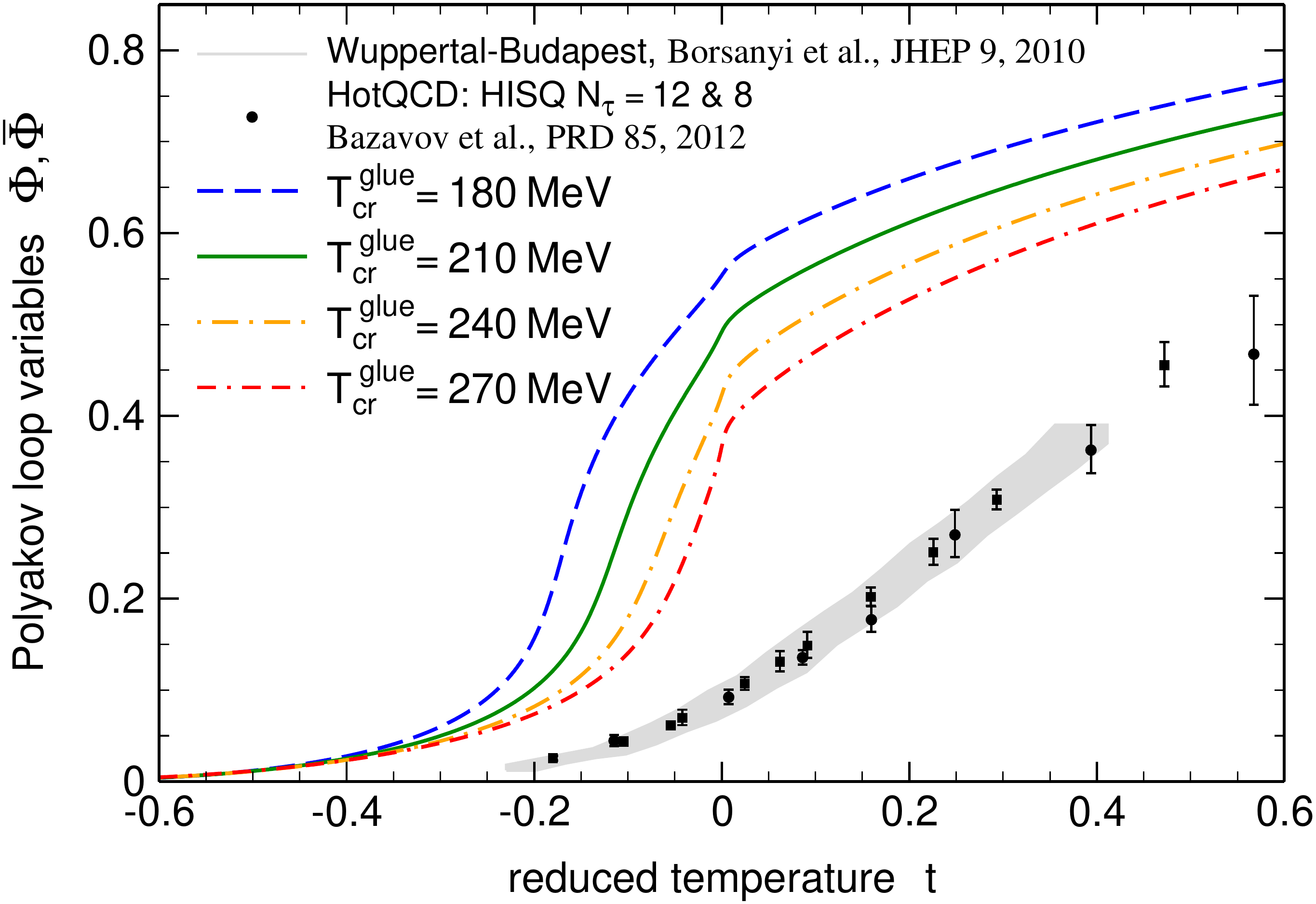}
	\caption[]{The Polyakov-loop expectation value $\Phi$ as
          function of temperature at $\mu_f=0$ with the adjustment to
          the glue potential for four different glue critical
          temperatures. The grey band is the continuum extrapolation
          of the lattice results of Ref.~\cite{Borsanyi:2010bp} and the
          data points lattice results of Ref.~\cite{Bazavov:2011nk}.}
	\label{fig:Ploop_T0s}
\end{figure}

In Fig.~\ref{fig:Ploop_T0s} we show the dependence of the evolution of
the Polyakov-loop expectation values on the critical temperature of
the glue Polyakov-loop potential.
The magnitude of the transition increases when the transition
temperature of the Polyakov-loop potential is lowered and the
transition regions is shifted towards smaller relative
temperatures. So the onset of deconfinement shifts towards the lower
end of the transition region with decreasing critical temperature of
the Polyakov-loop potential.  This is a natural consequence, as
decreasing the scale of the Polyakov-loop potential leads to a
relative shift of the glue part towards a lower scale in the full
theory.  Overall, the value of the Polyakov-loop variables shows a
large sensitivity on the critical temperature of the Polyakov-loop
potential but the general discrepancy to the lattice results is too
large for any constraints.

\begin{figure}
	\centering
	\includegraphics[width=0.47\textwidth]{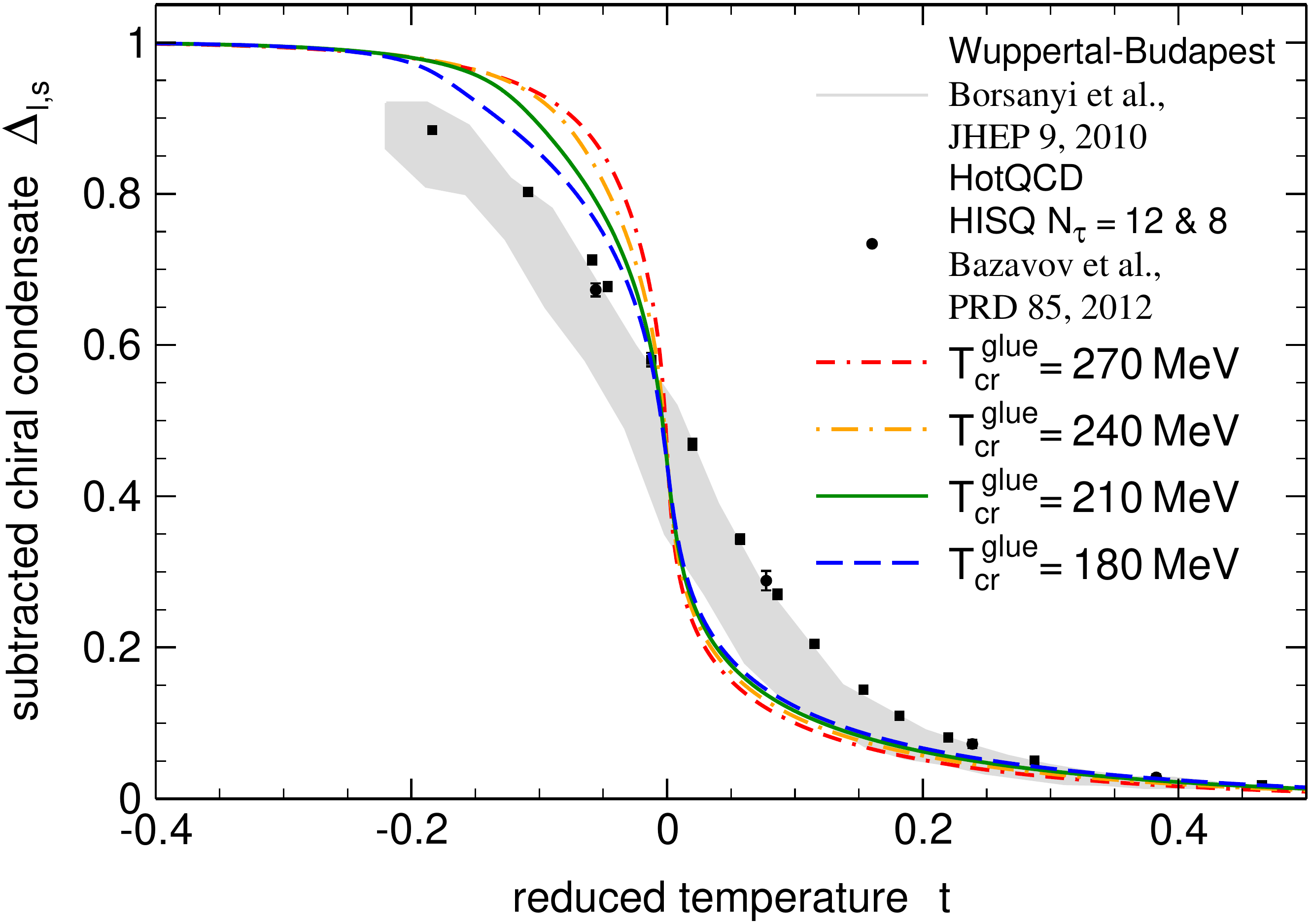}
	\caption[]{The subtracted chiral condensate $\Delta_\mrm{l,s}$
          as a function of temperature at $\mu_f=0$ with the
          adjustment to the glue potential for four different glue
          critical temperatures. The grey band is the continuum
          extrapolation of the lattice results of
          Ref.~\cite{Borsanyi:2010bp} and the data points lattice
          results of Ref.~\cite{Bazavov:2011nk}.}
	\label{fig:subcond_T0s}
\end{figure}

Due to the nontrivial coupling between Polyakov loop, quarks and
mesons, the onset of deconfinement at lower temperatures with
decreasing critical temperatures of the Polyakov-loop potential shows
its impact also in the order parameter for chiral symmetry, the
subtracted chiral condensate which is plotted in
Fig.~\ref{fig:subcond_T0s}.
It deviates at relative smaller
temperatures from its vacuum value when the transition temperature of
the Polyakov-loop potential is lowered. This effect gets enhanced due
to the simultaneous lowering of the pseudocritical temperature.
Apart from that, the chiral condensate shows only a very mild
dependence on the transition temperature of the glue potential.

\subsection{\label{ssec:ResPardep}Dependence on the parametrisation\protect}

The second uncertainty in the gauge sector of PNJL \& PQM models
besides the transition temperature of the Polyakov-loop potential is
its form of parametrisation.  We demonstrated in
Figs.~\ref{fig:PloopYM} and \ref{fig:e3pYM} that the different forms
and parameter sets show a significant difference already in pure YM
theory.  Ideally, the Polyakov-loop potential of full QCD from first
principal calculations should be applied. Since this is not yet
directly applicable one can at least consider the uncertainty that
comes along with the different parametrisations and parameter sets.

For the comparison of the different parametrisations of the
Polyakov-loop potential we choose a sigma meson mass of
$500\,\mrm{MeV}$ and a transition temperature of the Polyakov-loop
potential of $210\,\mrm{MeV}$.  For better readability of the plots we
omit in the following figures the results with the pure YM
Polyakov-loop potential and choose the same range of the ordinates as
in Sec.~\ref{ssec:Res_ImPoLoPot} to give an estimate of the reduction
due to the application of the quark-improved Polyakov-loop
potential. The figures with the results with both potentials will be
presented in Ref.~\cite{Stiele:2013gra}.

\begin{figure}
	\centering
	\includegraphics[width=0.47\textwidth]{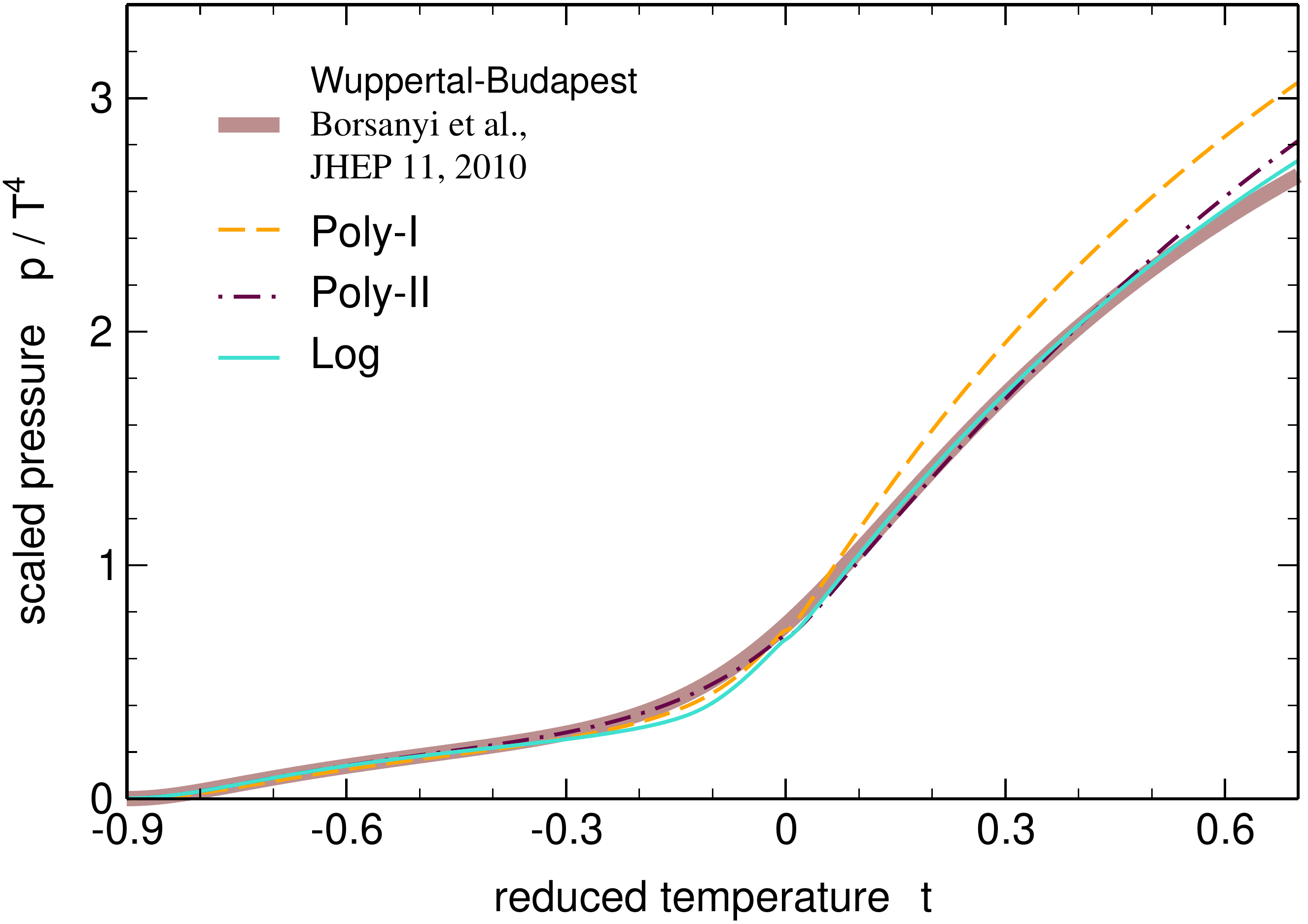}
	\caption[]{Scaled pressure $p/T^4$ as a function of
          temperature at $\mu_f=0$ with the adjustment to the glue
          potential for three different parametrisations and parameter
          sets of the Polyakov-loop potential. The orange band is the
          fit to the continuum extrapolation of the lattice results of
          Ref.~\cite{Borsanyi:2010cj}.}
	\label{fig:Pressure_Params}
\end{figure}

The normalised pressure (Fig.~\ref{fig:Pressure_Params}) shows the
steepest rise in the transition region for the polynomial
parametrisation with the parameters of Ref.~\cite{Scavenius:2002ru}
(\mbox{Poly-I} potential). This is in accord with the description of
pure Yang-Mills theory discussed in Sec.~\ref{ssec:PloopPot}.
Overall, the difference in the results for the scaled pressure for the
different parametrisations is not large around the transition region.

\begin{figure}
	\centering
	\includegraphics[width=0.47\textwidth]{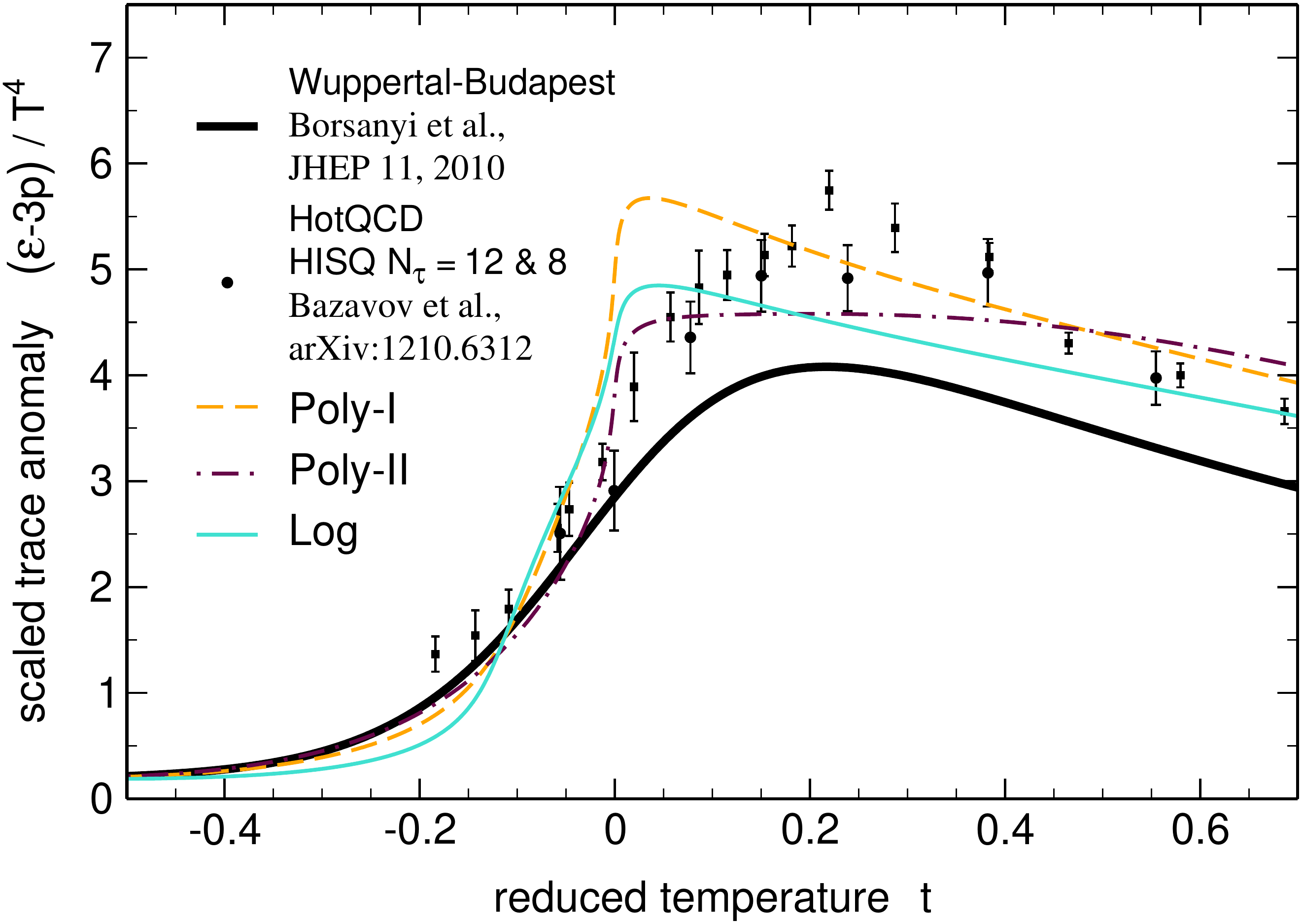}
	\caption[]{Scaled trace anomaly $\blr{\epsilon-3p}/T^4$ as a
          function of temperature at $\mu_f=0$ with the adjustment to
          the glue potential for three different parametrisations and
          parameter sets of the Polyakov-loop potential. The black
          band is the fit to the continuum extrapolation of the
          lattice results of Ref.~\cite{Borsanyi:2010cj} and the data
          points lattice results of the HotQCD
          collaboration~\cite{Bazavov:2012bp}.}
	\label{fig:e3p_Params}
\end{figure}

As seen before, the differences are more pronounced for the trace
anomaly or interaction measure (Fig.~\ref{fig:e3p_Params}). The slopes
of the curves are similar for the Poly-I and logarithmic
parametrisation but with a larger amplitude in the case of the Poly-I
potential. This originates from the description of pure Yang-Mills theory as
seen in Fig.~\ref{fig:e3pYM}.
The normalised trace anomaly with the Poly-II parametrisation of
Ref.~\cite{Ratti:2005jh} is nearly constant above the transition
temperature. This is a feature of the offset of the maximum of the
trace anomaly in Yang-Mills theory seen in Fig.~\ref{fig:e3pYM}. This offset
is also seen in calculations of full QCD with the Yang-Mills Polyakov-loop
potential \cite{Schaefer:2009ui}. The overall reduction of the amplitude
in the transition region for the improved glue potential leads to the
behaviour seen in Fig.~\ref{fig:e3p_Params}.

\begin{figure}[b]
	\centering
	\includegraphics[width=0.47\textwidth]{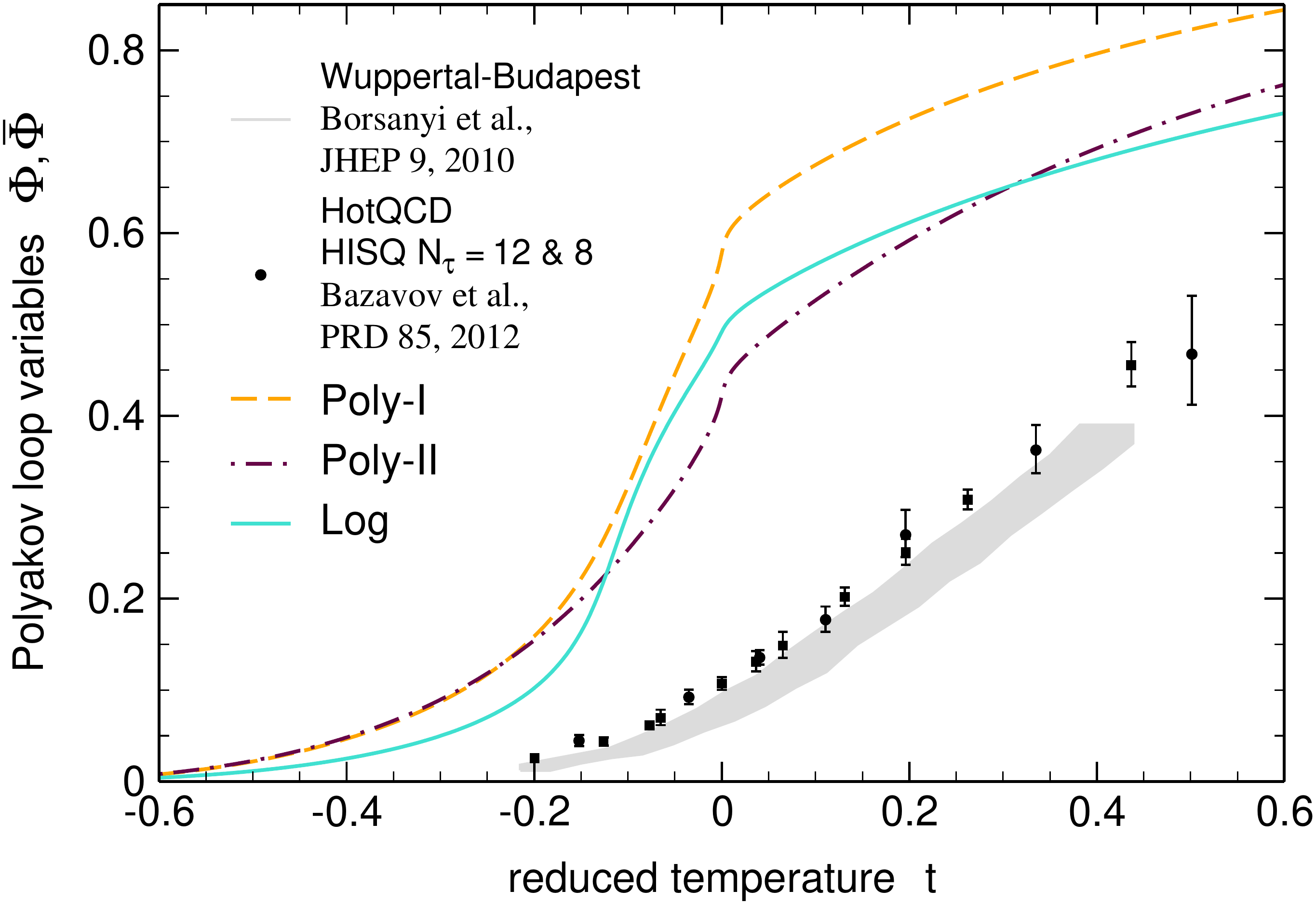}
	\caption[]{The Polyakov-loop expectation value $\Phi$ as
          function of temperature at $\mu_f=0$ with the adjustment to
          the glue potential for three different parametrisations and
          parameter sets of the Polyakov-loop potential. The grey band
          is the continuum extrapolation of the lattice results of
          Ref.~\cite{Borsanyi:2010bp} and the data points lattice
          results of Ref.~\cite{Bazavov:2011nk}.}
	\label{fig:Ploop_Params}
\end{figure}

The evolution of the Polyakov-loop variables shown in
Fig.~\ref{fig:Ploop_Params} reveals that the onset of the transition
in the confined phase is the steepest with the logarithmic
potential. The Poly-II parametrisation leads to the smoothest
evolution of the Polyakov loop below the pseudocritical temperature.
Note that we normalised the Polyakov-loop expectation value shown in
Fig.~\ref{fig:Ploop_Params} for the calculations with the polynomial
parametrisation such that it tends towards unity at infinite
temperature.  Overall, the differences in the Polyakov-loop variables
due to the different parametrisations are so large that it should be
considered before making quantitative statements.

\begin{figure}
	\centering
	\includegraphics[width=0.47\textwidth]{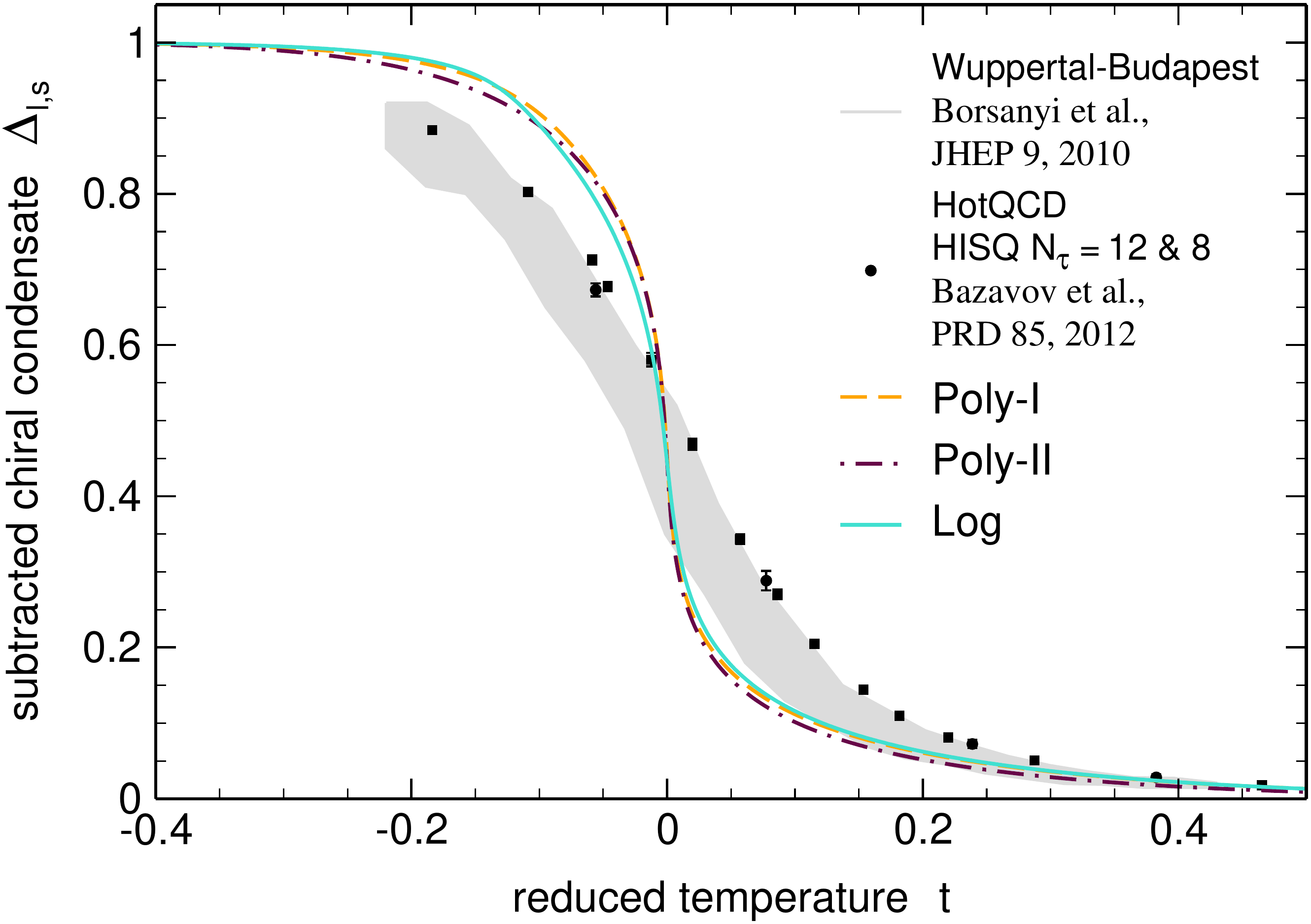}
	\caption[]{The subtracted chiral condensate $\Delta_\mrm{l,s}$
          as a function of temperature at $\mu_f=0$ with the
          adjustment to the glue potential for three different
          parametrisations and parameter sets of the Polyakov-loop
          potential. The grey band is the continuum extrapolation of
          the lattice results of Ref.~\cite{Borsanyi:2010bp} and the
          data points lattice results of Ref.~\cite{Bazavov:2011nk}, see also Ref.~\cite{Fischer:2012vc}.}
	\label{fig:DeltaLS_Params}
\end{figure}

Nevertheless, the differences in the gauge sector due to the different
parametrisations are considerably smaller in the evolution of the
subtracted chiral condensate (Fig.~\ref{fig:DeltaLS_Params}).
The slightly stronger transition in the gauge sector for the Poly-I
potential shifts the pseudocritical temperature to a smaller value
compared to the other cases as seen in Table
\ref{tab:critTemps_Params}.

\begin{table}
  \caption{Pseudocritical temperatures for the crossover transition 
    at $\mu_f=0$ for different parametrisations and parameter sets of 
    the Polyakov-loop potential. They are determined by the peaks of 
    the chiral susceptibility $\pd{\Delta_\mrm{l,s}}/\pd{T}$.}
	\begin{ruledtabular}
		\begin{tabular}{l|c|c|c}
						& Poly-I	& Poly-II	& Log \\ \hline
			$T_\mrm{c}$ [MeV]	& 144 	& 158 	& 158
		\end{tabular}
	\end{ruledtabular}
	\label{tab:critTemps_Params}
\end{table}

\subsection{\label{ssec:Resmsdep}Dependence on the sigma meson mass\protect}

Usually, the chiral partner of the pion is associated with the
resonance $f_0(500)$ \cite{Beringer:1900zz}. Note however that in
Ref.~\cite{Parganlija:2012fy} it was demonstrated that within an extended
quark-meson model that includes vector and axial-vector mesons the
resonance $f_0(1370)$ was identified as the nonstrange scalar
quarkonium state. Within our simple quark-meson model we consider
$m_\sigma=(400-600)\,\mrm{MeV}$ to be a reasonable parameter range.

To investigate the role of the mass of the scalar sigma meson the
logarithmic parametrisation of the Polyakov-loop potential with a
transition temperature of $210\,\mrm{MeV}$ is used in the following.

\begin{figure}
	\centering
	\includegraphics[width=0.47\textwidth]{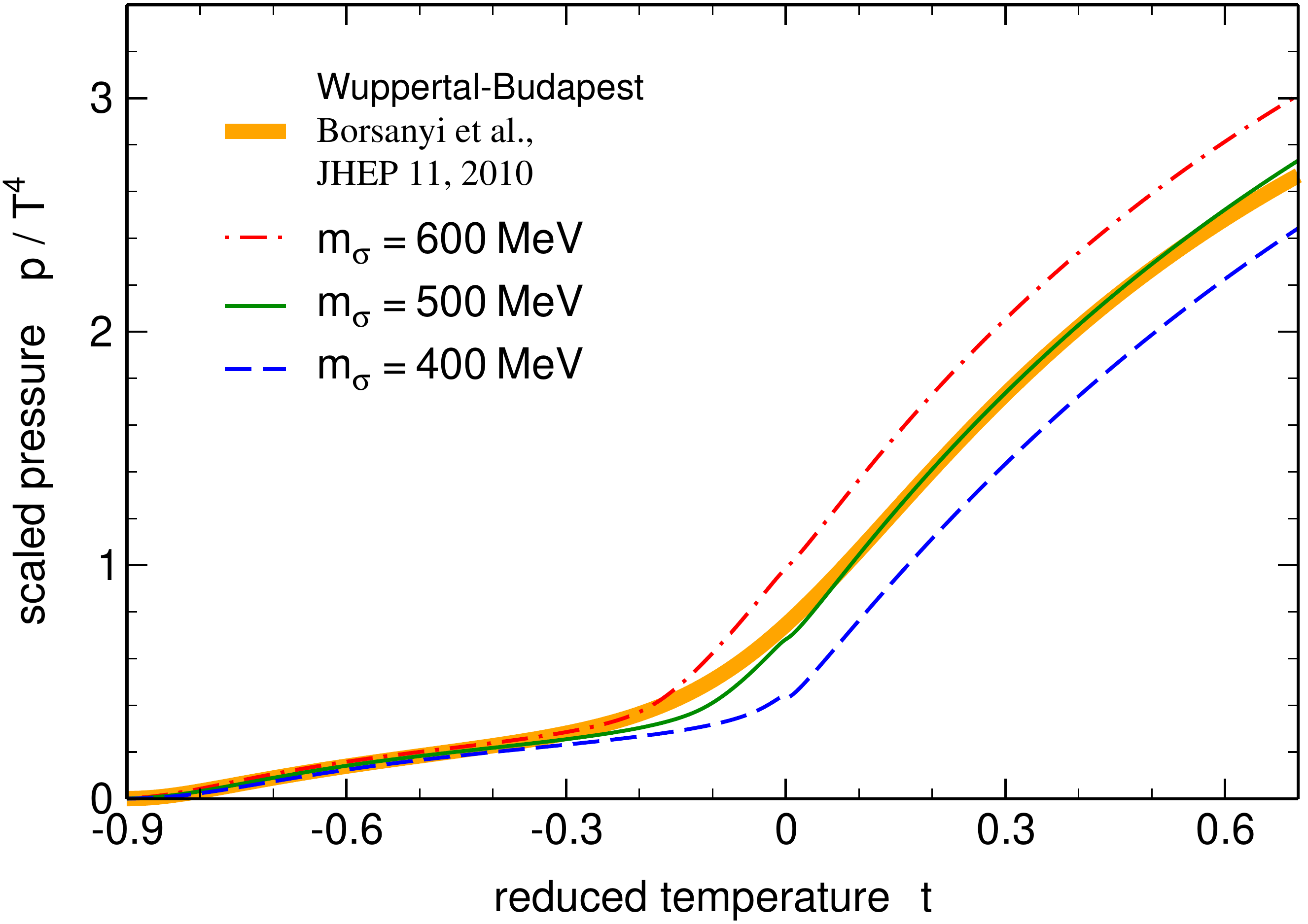}
	\caption[]{Scaled pressure $p/T^4$ as a function of
          temperature at $\mu_f=0$ with the adjustment to the glue
          potential for three different masses of the sigma meson. The
          orange band is the fit to the continuum extrapolation of the
          lattice results of Ref.~\cite{Borsanyi:2010cj}.}
	\label{fig:Pressure_msigma}
\end{figure}

The normalised pressure as displayed in Fig.~\ref{fig:Pressure_msigma}
shows that the steepness of the transition is relatively independent
on the mass of the $\sigma$-meson.
The location of the pseudocritical temperature relative to the
transition region leads to differences in the evolution. For $m_\sigma
= 400\,\mrm{MeV}$ the pseudocritical temperature is small (see Table
\ref{tab:critTemps_msigma}) and at the lower end of the transition
region so that the pressure just starts to rise at $T_\mrm{c}$.
\begin{table}
  \caption{Pseudocritical temperatures for the crossover transition at $\mu_f=0$ 
    for different masses of the sigma meson. They are determined by the peaks
    in the temperature derivatives of the subtracted condensate $\Delta_\mrm{l,s}$ .}
	\begin{ruledtabular}
		\begin{tabular}{l|c|c|c}
			$m_\sigma$ [MeV]	& 400 & 500 & 600 \\ \hline
			$T_\mrm{c}$ [MeV] 	& 144 & 158 & 173
		\end{tabular}
	\end{ruledtabular}
	\label{tab:critTemps_msigma}
\end{table}
The evolution for $m_\sigma=600\,\mrm{MeV}$ is the other
extreme. \pagebreak Here, the scaled pressure rises already significantly below
$T_\mrm{c}$ since the pseudocritical temperature is just reached at
the upper end of the transition region and at a relatively large
absolute value as given in Table \ref{tab:critTemps_msigma}.

The change of the pressure for an increasing mass of the sigma meson
(Fig.~\ref{fig:Pressure_msigma}) is opposite to the case of an
increasing glue critical temperature
(Fig.~\ref{fig:Pressure_T0s}). Therefore, to get agreement with the
pressure of the lattice calculation for a sigma meson mass of
$400\,\mrm{MeV}$ one would have to choose a critical temperature of
the glue potential of $180\,\mrm{MeV}$ and for
$m_\sigma=600\,\mrm{MeV}$ one would require
$T^{\mrm{glue}}_{\mrm{cr}}=250\,\mrm{MeV}$. So the lower the mass of
the sigma meson, the smaller the transition temperature of the glue
potential has to be in order to reproduce the lattice results.  This
is due to the fact that a decrease in the mass of the sigma meson
lowers the scale of the chiral transition as can be seen in Table
\ref{tab:critTemps_msigma} and Ref.~\cite{Schaefer:2008hk}. So the scale
of the (de)confinement transition has to decrease as well.  Note that
the combination of $m_\sigma = 500\,\mrm{MeV}$ and
$T^{\mrm{glue}}_{\mrm{cr}}=210\,\mrm{MeV}$ leads to a pseudocritical
temperature that is equal to the one of lattice calculations.

\begin{figure}
	\centering
	\includegraphics[width=0.47\textwidth]{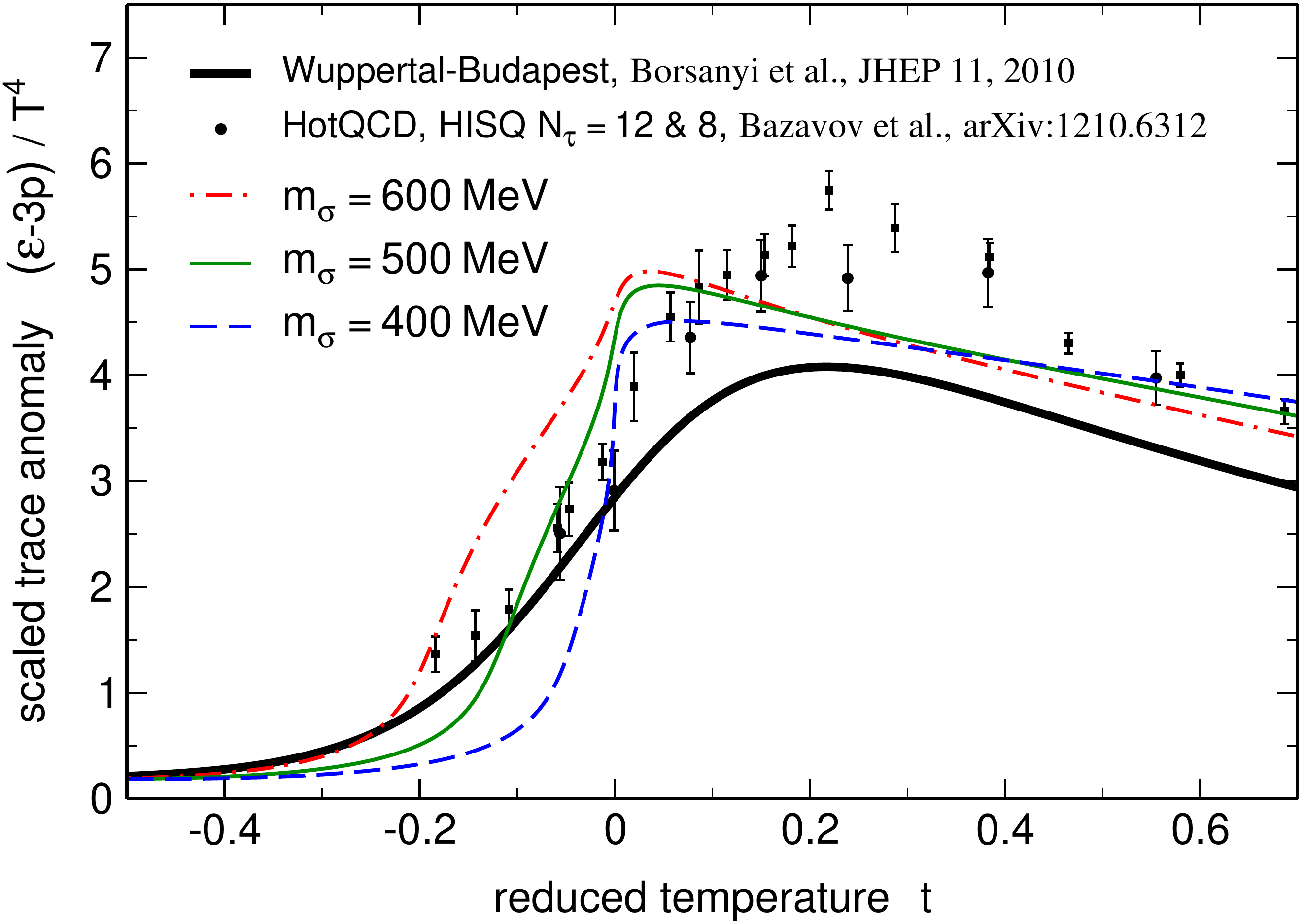}
	\caption[]{Scaled trace anomaly $\blr{\epsilon-3p}/T^4$ as a
          function of temperature at $\mu_f=0$ with the adjustment to
          the glue potential for three different masses of the sigma
          meson. The black band is the fit to the continuum
          extrapolation of the lattice results of
          Ref.~\cite{Borsanyi:2010cj} and the data points lattice
          results of the HotQCD collaboration~\cite{Bazavov:2012bp}.}
	\label{fig:e3p_msigmas}
\end{figure}

Figure \ref{fig:e3p_msigmas} shows the results of the trace
anomaly for different sigma meson masses.
The transition region broadens for a larger mass of the $\sigma$-meson.

\begin{figure}
	\centering
	\includegraphics[width=0.47\textwidth]{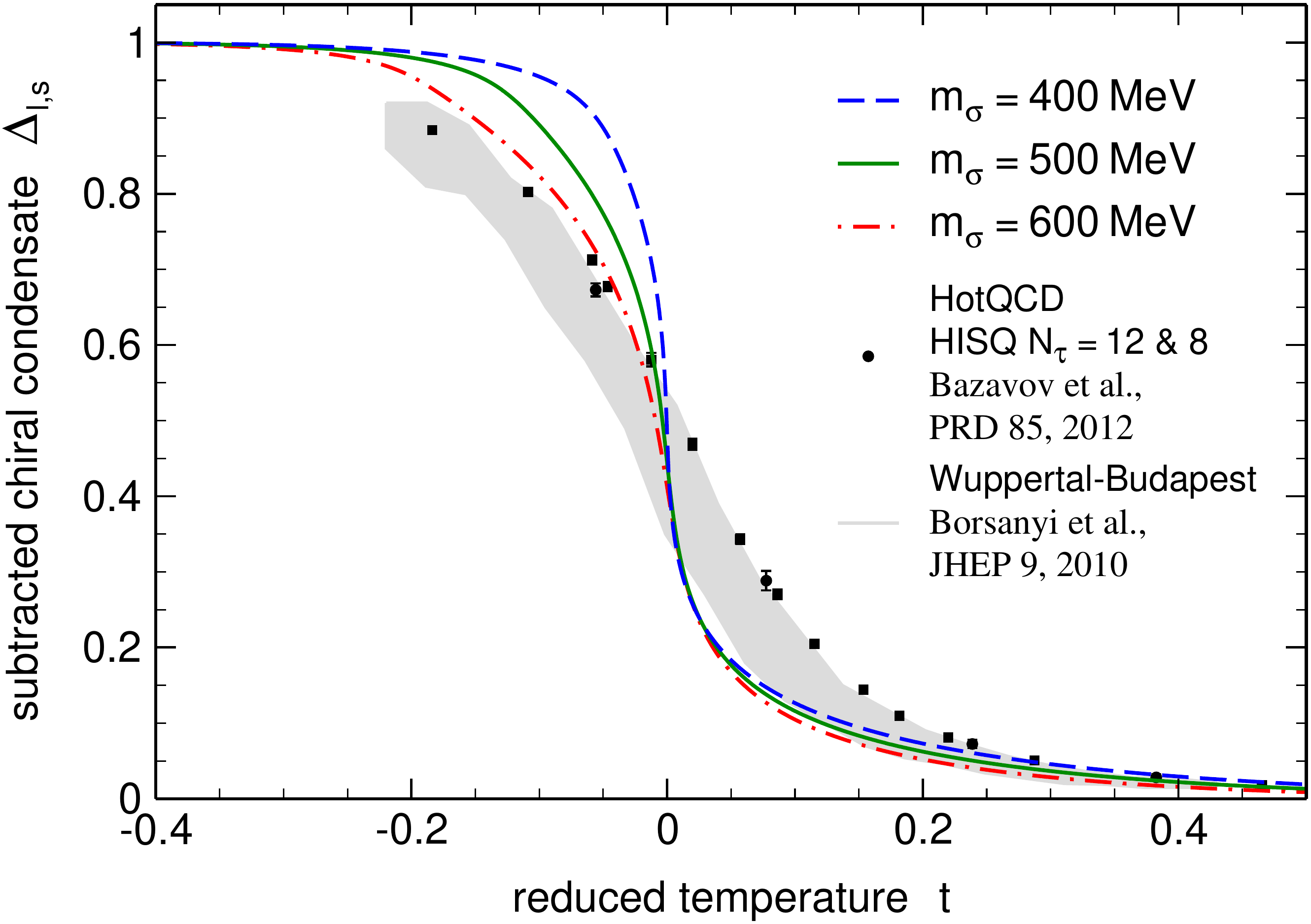}
	\caption[]{The subtracted chiral condensate $\Delta_\mrm{l,s}$
          as a function of temperature at $\mu_f=0$ with the
          adjustment to the glue potential for three different masses
          of the sigma meson. The grey band is the continuum
          extrapolation of the lattice results of
          Ref.~\cite{Borsanyi:2010bp} and the data points lattice
          results of Ref.~\cite{Bazavov:2011nk}.}
	\label{fig:DeltaLS_msigma}
\end{figure}

The subtracted condensate is shown in Fig.~\ref{fig:DeltaLS_msigma}.
The larger the mass of the sigma meson, the earlier the chiral
condensate deviates from its vacuum expectation value. The temperature
dependence of the chiral condensate becomes smoother with larger
$\sigma$-meson masses.

\begin{figure}
	\centering
	\includegraphics[width=0.47\textwidth]{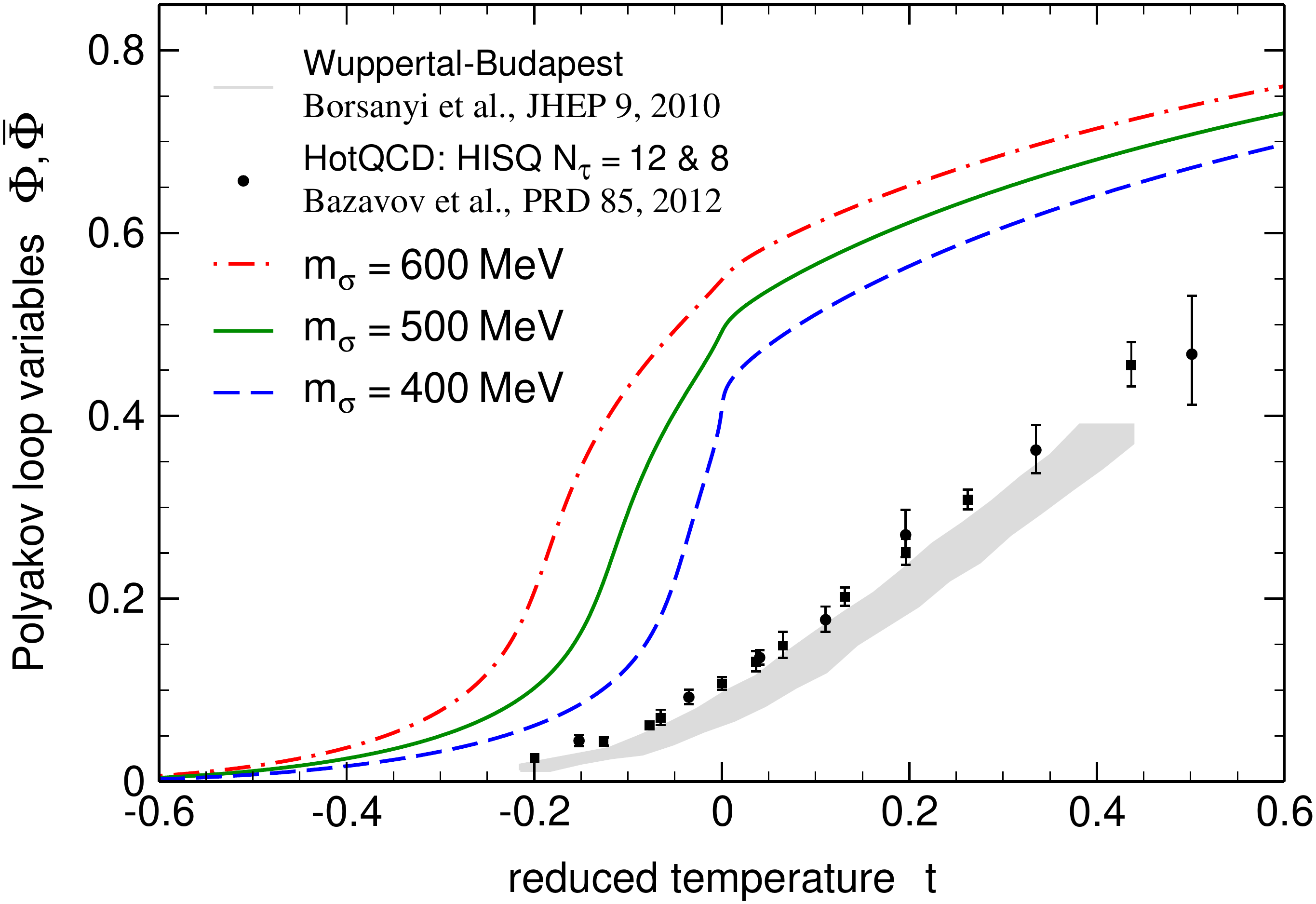}
	\caption[]{The Polyakov-loop expectation value $\Phi$ as
          function of temperature at $\mu_f=0$ with the adjustment to
          the glue potential for three different masses of the sigma
          meson. The grey band is the continuum extrapolation of the
          lattice results of Ref.~\cite{Borsanyi:2010bp} and the data
          points lattice results of Ref.~\cite{Bazavov:2011nk}.}
	\label{fig:Ploop_msigma}
\end{figure}

The evolution of the Polyakov-loop variables
(Fig.~\ref{fig:Ploop_msigma}) shows that the transition regions
shrinks when the mass of the sigma meson is lowered. Decreasing the
glue critical temperature broadens the transition region towards lower
temperatures. The pseudocritical temperatures for
$m_\sigma=400\,\mrm{MeV}$ is already smaller than the lattice result
for $T^{\mrm{glue}}_{\mrm{cr}}=210\,\mrm{MeV}$.

\pagebreak
\section{\label{sec:DiscOut}Conclusions \& Outlook\protect}

In this work we have presented results for the nonperturbative
Polyakov-loop potential in QCD in terms of the expectation value
$\langle A_0\rangle$. The potential $V[\langle A_0\rangle]$ is an
order-parameter potential for the deconfinement transition. It is
obtained from the first-principle QCD flows in
Refs.~\cite{Braun:2009gm,Pawlowski:2010ht} solely from the QCD two-point
functions of the ghost and the gluon fields. In particular this
computation takes into account the full backreaction of the quarks
and the nonlinear coupling of the quark-gluon sector. So far, this
important physics has been omitted in Polyakov-loop potentials used in
PNJL/PQM models, where a Polyakov-potential is used which
is generated solely from lattice Yang-Mills data for the Polyakov loop and
the thermodynamic observables.

In Sec~\ref{sec:FRG}, we have compared the glue Polyakov loop
potential with that in pure Yang-Mills theory
\cite{Braun:2007bx,Braun:2010cy,Fister:2013bh}.  We have shown that
the two potentials are simply related by an appropriate rescaling of
the temperature, see Eq.~\eq{eq:gluePotYMPot}. Such an approach allows
to mimic the effect of the quark backreaction on the gauge sector
describing the confining dynamics.  Whereas the fit of the parameters
of the Polyakov-loop potential to the lattice data contains only
information about the minimum of the potential in the deconfined
phase, our fit also adjusts the potential in the confined phase. 
In total this provides a simple method for improving effective
low-energy models towards full QCD.

In the second part of this work, we have applied this improvement in the
2+1 flavour PQM model. Our adjustment of the temperature scale of the
Polyakov-loop potentials leads to a softening of the temperature
dependence of physical observables at temperatures close to the chiral
and deconfinement crossover at vanishing chemical potential. As a
consequence, the behaviour of the order parameters and thermodynamic
observables is in much better agreement with latest results from
lattice QCD studies and first-principles functional
approaches. Specifically, we find good agreement with the scaled
pressure and the trace anomaly from lattice QCD simulations, even for
the amplitude of the trace anomaly.

In order to test the robustness of our results, we have analysed the
dependence on the parametrisation and the transition temperature of
the Polyakov-loop potential as well as the $\sigma$-meson mass.  We
have found a parametrisation-dependence of the results, which can not
be ignored in a quantitative analysis. To be more specific, we have
varied the glue critical temperature and the mass of the scalar
$\sigma$-meson and observed that a larger glue critical temperature
has a similar effect on our results as a decrease of the mass of the
$\sigma$-meson. In comparison to the computation without QCD-improved
Polyakov-loop potential we find a qualitatively better agreement with
the lattice data, see in particular the trace anomaly in
\Fig{fig:TraceAnomaly}. For a sigma meson mass $m_{\sigma} \approx
500\,\text{MeV}$ and a transition temperature of the glue potential of
$\sim 210~\text{MeV}$ this agreement is most pronounced.

We have also discussed the next steps for systematically improving
PNJL/PQM models towards low-energy QCD. These steps include the
determination of the input parameters in the chiral sector,
i.e., $m_{\sigma}$, from the QCD flows. Furthermore, the model
potentials can be upgraded by successively taking into account more
information from the full glue potential. One possibility is e.g.\ to
work-in the correct height of the barrier and the difference of the
potential at the minima in the metastable region. This has a direct
impact on the steepness of both transitions. Finally the QCD potential
obtained from the flow equation \cite{Braun:2009gm,Pawlowski:2010ht}
can be used directly, see also
Refs.~\cite{Braun:2011fw,Braun:2012zq,Fukushima:2012qa}. Moreover, matter
fluctuations can be included within an extension
of the present work and Ref.~\cite{Herbst:2013ail}. 

In conclusion, the present work offers a simple and systematic
approach for improving the gauge sector of low-energy effective models
towards QCD. The comparison of our results with that from lattice QCD
confirms that this is important as it significantly affects the
behaviour of physical observables in the crossover region.

\begin{acknowledgments}
	We thank L.~Fister, E.~S.~Fraga, T.~K.~Herbst, B.~W.~Mintz and B.-J.~Schaefer for
	discussions and collaboration on related topics. This work is
	supported by BMBF under grants FKZ 05P12VHCTG and 06HD7142, by
	ERC-AdG-290623, by the Helmholtz Alliance HA216/EMMI, by the DFG
	through the HGSFP,  by HGS-HIRe, and by HIC for FAIR within the LOEWE program.
\end{acknowledgments}

\bibliography{refs}

\end{document}